\documentclass[journal]{IEEEtran}

\usepackage{cite}
\usepackage{hyperref}

\ifCLASSINFOpdf
    \usepackage[pdftex]{graphicx}
    \graphicspath{{./figures/}}
\else
\fi

\usepackage{mathtools}
\usepackage{algorithm}
\usepackage{algorithmic}
\usepackage{eqparbox}

\usepackage{xcolor}
\usepackage{xspace} 

\ifCLASSOPTIONcompsoc
    \usepackage[caption=false,font=normalsize,labelfont=sf,textfont=sf]{subfig}
\else
  \usepackage[caption=false,font=footnotesize]{subfig}
\fi
\usepackage{siunitx}
\usepackage{mathptmx}
\usepackage{threeparttable}
\usepackage{amsmath,amssymb,amsfonts}
\usepackage{algorithmic}
\usepackage{textcomp}
\usepackage{array}
\usepackage{multirow}

\begin{document}
\bstctlcite{IEEEexample:BSTcontrol} 

\title{Compressive Sensing Photoacoustic Imaging Receiver with Matrix-Vector-Multiplication SAR ADC}

\author{
        Huan-Cheng~Liao,~\IEEEmembership{Student Member,~IEEE,}
        Shunyao~Zhang,
        Yumin~Su,~\IEEEmembership{Student Member,~IEEE,}
        Arvind~Govinday,
        Yiwei~Zou,~\IEEEmembership{Student Member,~IEEE,}
        Wei~Wang,~\IEEEmembership{Student Member,~IEEE,}
        Vivek~Boominathan,~\IEEEmembership{Member,~IEEE,}
        Ashok~Veeraraghavan,~\IEEEmembership{Fellow,~IEEE,}
        Lei~S.~Li,
        and~Kaiyuan~Yang,~\IEEEmembership{Senior Member,~IEEE}
\thanks{Manuscript received on ... This work is supported in part by the National Science Foundation (NSF) under awards 2443135 and 2023849 \textit{(Huan-Cheng Liao and Shunyao Zhang contributed equally to this work.) (Corresponding author: Kaiyuan Yang)}}
\thanks{The authors are with the Department of Electrical and Computer Engineering, Rice University, Houston TX, 77005, USA. (e-mail: kyang@rice.edu)}
}

\markboth{Journal of Solid-State Circuits}%
{Shell \MakeLowercase{\textit{et al.}}: A Sample Article Using IEEEtran.cls for IEEE Journals}


\maketitle

\begin{abstract}
Wearable photoacoustic imaging devices hold great promise for continuous health monitoring and point-of-care diagnostics. However, the large data volume generated by high-density transducer arrays presents a major challenge for realizing compact and power-efficient wearable systems. This paper presents a photoacoustic imaging receiver (RX) that embeds compressive sensing directly into the hardware to address this bottleneck. The RX integrates 16 AFEs and four matrix-vector-multiplication (MVM) SAR ADCs that perform energy- and area-efficient analog-domain compression. The architecture achieves a 4–8x reduction in output data rate while preserving low-loss full-array information. The MVM SAR ADC executes passive and accurate MVM using user-defined programmable ternary weights. Two signal reconstruction methods are implemented: (1) an optimization approach using the fast iterative shrinkage-thresholding algorithm, and (2) a learning-based approach employing implicit neural representation. Fabricated in 65 nm CMOS, the chip achieves an ADC's SNDR of 57.5 dB at 20.41 MS/s, with an AFE input-referred noise of 3.5 nV/$\mathbf{\sqrt{Hz}}$. MVM linearity measurements show $\mathbf{R^2>0.999}$ across a wide range of weights and input amplitudes. The system is validated through phantom imaging experiments, demonstrating high-fidelity image reconstruction under up to 8x compression. The RX consumes 5.83 mW/channel and supports a general ternary-weighted measurement matrix, offering a compelling solution for next-generation miniaturized, wearable PA imaging systems.
\end{abstract}


\section{Introduction}
\IEEEPARstart{D}{eep-tissue} imaging is a vital tool in healthcare and medicine, offering noninvasive access to internal structures and biomarkers beneath the skin. This capability allows for early detection and continuous monitoring of a wide range of health conditions, including but not limited to musculoskeletal \cite{protopappas_ultrasound_2005}, fetal development \cite{ryu_comprehensive_2021}, and cardiovascular diseases \cite{hu_wearable_2023}. By aiding clinicians to identify abnormalities at earlier stages, deep-tissue imaging facilitates more effective diagnosis and treatment.
Deep tissue imaging can be broadly classified into two categories based on the imaging modality. The first is optical imaging \cite{stuker_fluorescence_2011}, \cite{kodach_quantitative_2010},\cite{miller2017deep}, which utilizes light to excite biological tissue and detects the resulting signal through an optical receiver (RX). This technique enables molecular contrast by exploiting wavelength-dependent absorption characteristics of specific biomolecules. However, its imaging depth is fundamentally limited due to strong light scattering in tissue. As light propagates, it undergoes multiple scattering, rapidly deviating from its original path and resulting in significant attenuation within a short distance, typically around 1 mm. 
\begin{figure}[t]
\centering
\includegraphics[width=.92\linewidth]{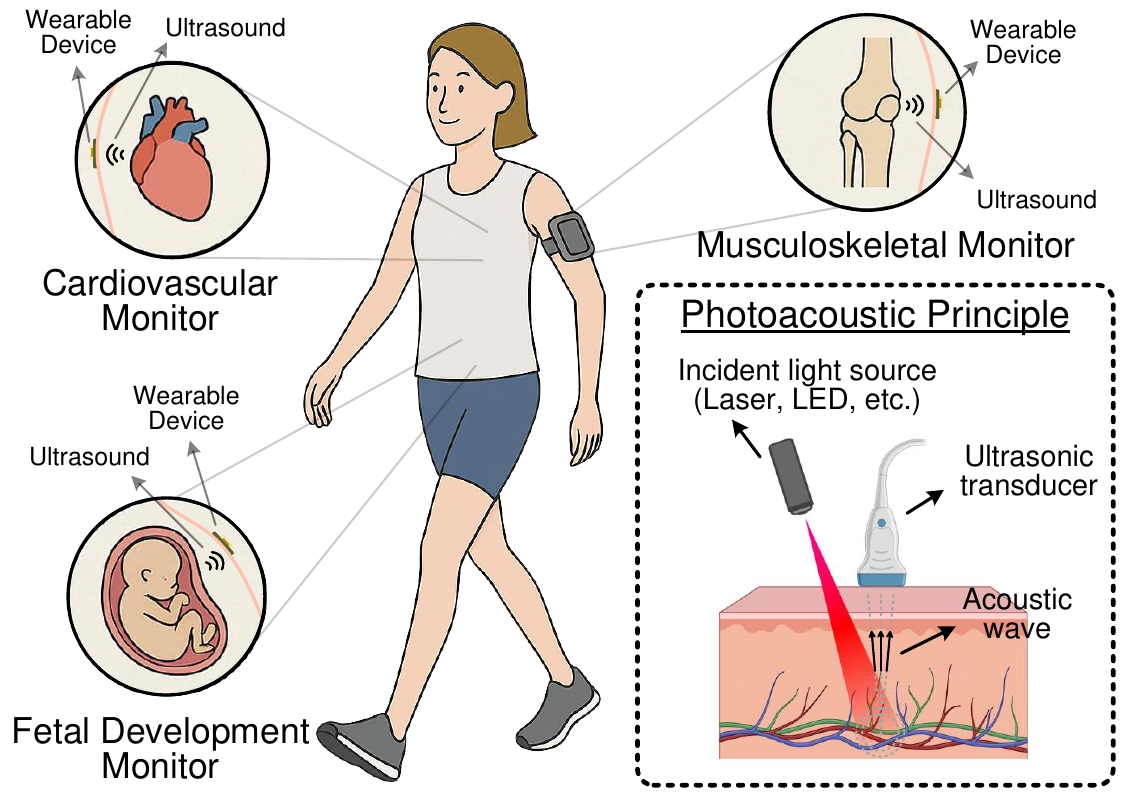}%
\vskip -2ex
\caption{Selected applications of wearable ultrasound/photoacoustic imaging devices and the principles of photoacoustic imaging.}
\label{fig:Fig1}
\end{figure}

The other modality is acoustic imaging. Ultrasound imaging \cite{chen_column-row-parallel_2016} uses acoustic waves for both transmission and reception, allowing the signals to penetrate much deeper into the tissue ($>$10 cm). While ultrasound imaging has already been broadly adopted in clinical settings, it does not provide molecular contrast.
Over the past decade, photoacoustic imaging, or photoacoustic tomography (PAT), has emerged as a promising hybrid modality that combines the molecular sensitivity of optical imaging with the deep tissue penetration of ultrasound \cite{wang_photoacoustic_2012, xu_photoacoustic_2006}. In PAT, the target tissue is illuminated by a pulsed light source, leading to localized absorption and transient heating as illustrated in Fig.~\ref{fig:Fig1}. The heat induces thermoelastic expansion and generates broadband acoustic waves, which are detected by an ultrasound RX. These acoustic signals encode spatial and molecular information about the absorbing structures and can be reconstructed into images through inverse algorithms. By leveraging optical contrast and acoustic propagation, photoacoustic imaging offers insights at depths beyond the reach of conventional optical methods and promises a powerful noninvasive tool for diagnosing various diseases, including lymphatic diseases, Crohn’s disease, skin disorders, cancers, and many others \cite{park_clinical_2024}, \cite{siphanto_serial_2005}.

Despite the impressive imaging capabilities, existing PAT systems are bulky benchtop platforms designed primarily for clinical or hospital use, such as the setup in \cite{kim_deeply_2010}. The large form factor limits the broader adoption of photoacoustic imaging, particularly in applications needing continuous monitoring and point-of-care accessibility. We envision a wearable photoacoustic imaging device that allows patients to conveniently and comfortably monitor their physiological conditions at home or in everyday settings, eliminating the need for frequent hospital visits. Such a system could provide real-time and long-term imaging data, offering insights into disease progression, treatment response, and overall health status.
However, translating photoacoustic imaging devices into a wearable form factor presents several critical challenges at the system and circuit levels. The device must be compact and power-efficient to support extended operation without compromising user comfort or mobility. Additionally, high-channel-count photoacoustic imaging systems typically generate massive volumes of data. For example, a 1024-channel array can easily generate over 300 Mbps of raw data, necessitating high-throughput wireless communication schemes, which often incur significant power overhead. To meet the stringent power and bandwidth constraints of wearable platforms, efficient data compression will be a key technology enabler for practical deployment.

\begin{figure}[t]
\centering
{\includegraphics[width=\linewidth]{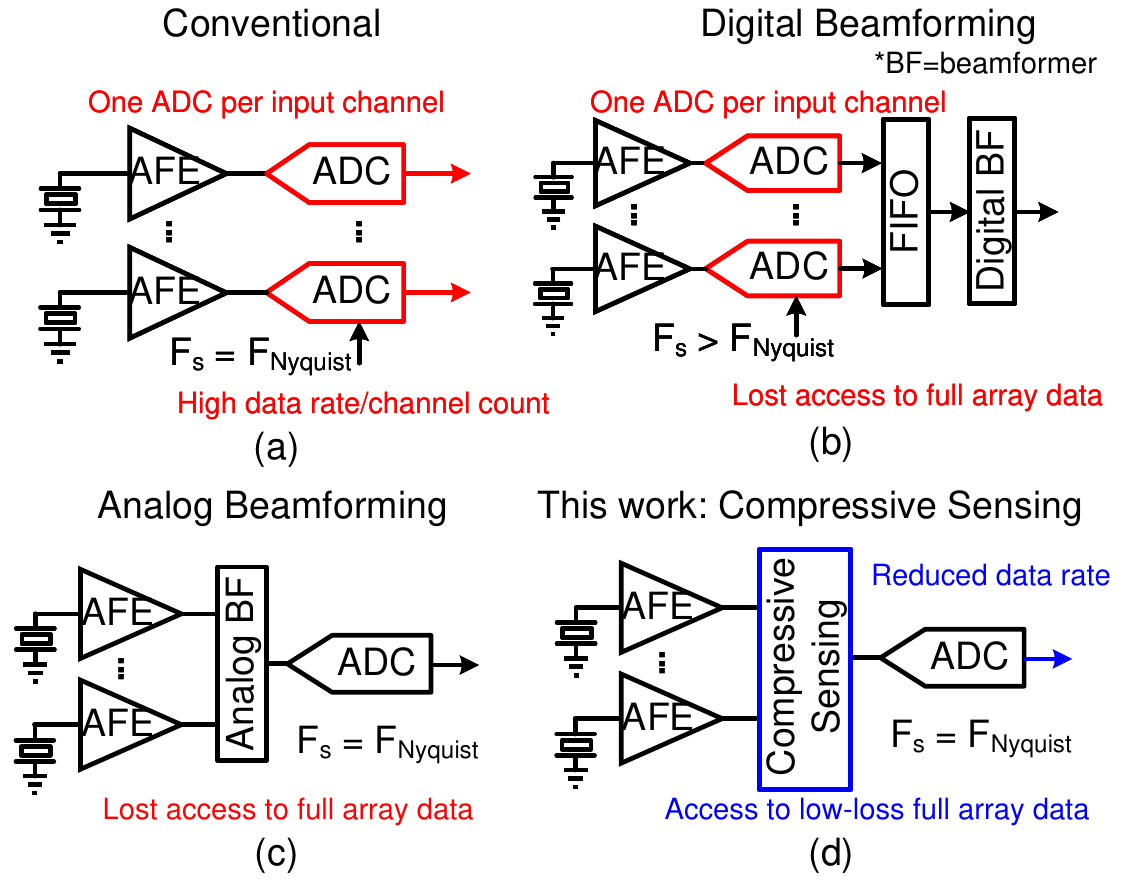}%
}
\vskip -2ex
\caption{Photoacoustic and ultrasound RX designs using (a) conventional architecture \cite{li_154mwelement_2019}, (b) digital beamforming \cite{chen_pixel_2017}, \cite{kim_single-chip_2017}, (c) analog beamforming \cite{guo_125_2024}, \cite{gurun_analog_2012}, and (d) compressive sensing.}
\label{fig:RX_prior_this}
\end{figure} 

Figure \ref{fig:RX_prior_this} summarizes the prior works on photoacoustic and ultrasound RXs, which share similar sensor arrays and readout circuits. The amplitude of received photoacoustic signals is generally an order of magnitude or more smaller than that of ultrasound echoes (on the order of MPa vs. $<$10 kPa)~\cite{beard2011biomedical}, depending on factors such as tissue absorption, transducer characteristics, and optical excitation conditions.
The most conventional approach, as shown in Fig.~\ref{fig:RX_prior_this}~a, employs one ADC per channel \cite{li_154mwelement_2019}. This architecture gives the backend reconstruction full access to raw data and places no limitations on image reconstruction. However, it results in a high output data rate and a high output channel count, which are unsuitable for compact and power-constrained systems. To address this limitation, on-chip beamforming have been implemented in the digital domain \cite{chen_pixel_2017}, \cite{kim_single-chip_2017}, as illustrated in Fig.~\ref{fig:RX_prior_this}~b, or analog domain \cite{guo_125_2024}, \cite{gurun_analog_2012}, as shown in Fig.~\ref{fig:RX_prior_this}~c. Beamforming reduces the number of output channels, and analog beamforming further decreases the number of required ADCs. Nevertheless, implementing beamforming in hardware limits the reconstruction side’s access to the full array data, incurring focusing errors and/or increased grating and sidelobe levels \cite{soozande_imaging_2022}, \cite{hopf_pitch-matched_2022}. While narrowing the transmit beams can help mitigate this issue, it compromises the frame rate. Therefore, a tradeoff exists between image quality and data rate in beamforming.

To break this tradeoff, this work exploits the compressive sensing technique in the RX to achieve full-array data acquisition at a reduced data rate with minimal information loss, as illustrated in Fig.~\ref{fig:RX_prior_this}~d. Compressive sensing has been successfully deployed in ultrasound imaging \cite{kruizinga_compressive_2017}, with the goal of enabling low-cost imaging with a single transducer. It relies on acquiring multiple measurements through mechanical rotation to produce sufficient information for image reconstruction, thus compromising temporal resolution. Meanwhile, the use of a fixed mask imposes limitations on post-fabrication flexibility, restricting its scalability to different imaging scenarios and reconstruction algorithms. In this work, by embedding compressive sensing directly into the RX, we present a system that preserves image quality while addressing stringent power and bandwidth constraints, achieving a 4–8x reduction in data rate and a 4x reduction in the number of required ADCs. The architecture employs analog-domain spatial compression using programmable ternary-weighted measurement matrices. It supports two reconstruction strategies: an optimization algorithm and a learning-based neural network approach. This design enables efficient, high-fidelity imaging, making it well-suited for wearable photoacoustic systems.

This article extends \cite{liao_352_2025} and is organized as follows: Section II presents the principles of compressive sensing and design considerations for photoacoustic imaging RX. Section III provides details on the implementation of the circuit and system designs. Section IV presents the experimental results, including chip measurement and imaging system. Section V concludes this article.
\section{Principles of Compressive Sensing Receiver}
Compressive sensing exploits signal sparsity to reduce the required number of samples, enabling lower data rates while maintaining acceptable reconstructed signal quality \cite{baraniuk_compressive_2007}. The following briefly overviews compressive sensing and discusses its implementation for photoacoustic imaging RX. Detailed explanations on the compressive sensing theory can be found in \cite{donoho_compressed_2006}, \cite{candes_sparsity_2007}.

\subsection{Compressive Sensing Theory}
Compressive sensing enables the recovery of sparse signals from fewer measurements. A discrete-time signal \(x\in \mathbb{R}^M\) is projected onto a lower-dimensional measurement space \(y\in \mathbb{R}^N \) using a measurement matrix \(\Phi\in\mathbb{R}^{N\times M} \), where \(N < M\). The measurement procedure is expressed as
\begin{equation}
y = \Phi x.
\end{equation}

Since \(N < M\), the output vector y has a lower dimensionality than the input x, thereby achieving data compression. Although this underdetermined system has infinitely many solutions, recovery is possible if \( x \) is sparse or compressible in a known basis \( \Psi \in \mathbb{R}^{M \times M} \), such that \( x = \Psi \alpha \), where \( \alpha \) contains only \( S < M \) non-zero coefficients. Substituting into the measurement equation yields

\begin{equation}
y = \Phi \Psi \alpha = \Theta \alpha,
\end{equation}

where \( \Theta = \Phi \Psi \in \mathbb{R}^{N \times M} \) is the effective measurement matrix.
Accurate recovery of \( \alpha \) from \( y \) requires that \( \Theta \) satisfy the Restricted Isometry Property (RIP) \cite{candes_decoding_2005}, which ensures that the geometry of all \( S \)-sparse signals is approximately preserved during projection. Specifically, for all \( S \)-sparse \( x \), the RIP condition is given by

\begin{equation}
(1 - \delta_S)\|x\|_2^2 \leq \|\Phi x\|_2^2 \leq (1 + \delta_S)\|x\|_2^2,
\end{equation}

where \( \delta_S\) is the isometry constant. Random matrices with i.i.d. Gaussian or Bernoulli-distributed entries are known to satisfy the RIP with high probability. Under these conditions, sparse signal recovery can be achieved via convex optimization or greedy algorithms. 

\begin{figure}[!t]
\centering
{\includegraphics[width=\linewidth]{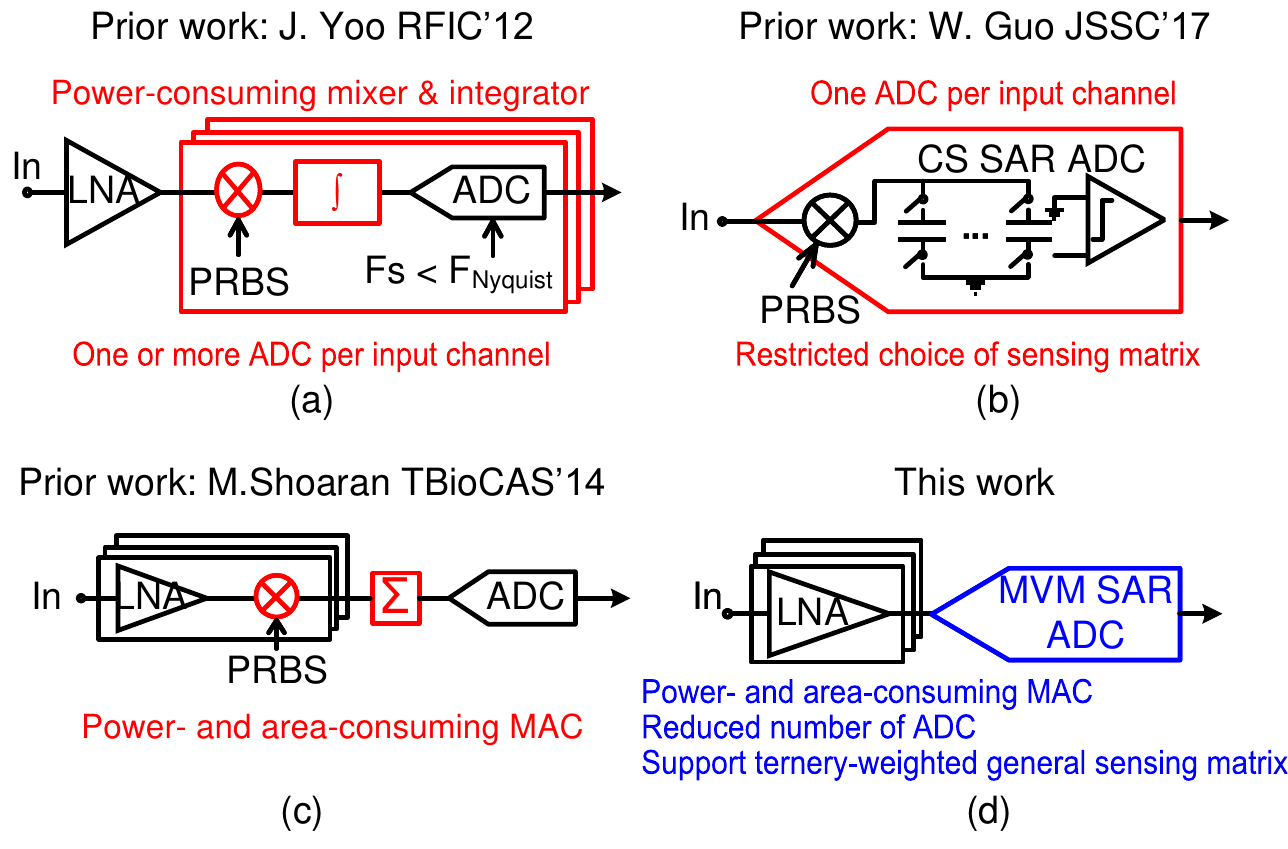}%
} 
\vskip -2ex
\caption{Compressive sensing RX designs based on (a) time-domain active MVM \cite{yoo_100mhz2ghz_2012}, (b) time-domain passive MVM \cite{guo_fully_2017}, (c) spatial-domain active MVM \cite{shoaran_compact_2014}, and (d) spatial-domain passive MVM.}
\label{fig:CS_prior_this}
\end{figure}

\subsection{Compressive Sensing Photoacoustic Imaging RX}
\subsubsection{Hardware Design Considerations}
Two primary design choices for implementing compressive sensing in the RX are time-domain compression and spatial-domain compression. As shown in Fig.~\ref{fig:CS_prior_this}~(a,b), prior works \cite{yoo_100mhz2ghz_2012}, \cite{guo_fully_2017} demonstrated that time-domain compression can be integrated into hardware design, which reduces the output data rate and lowers the ADC sampling frequency requirement. However, the design in \cite{yoo_100mhz2ghz_2012} relies on power-consuming active blocks such as mixers and integrators to implement the matrix-vector-multiplication (MVM), which is the core operation of the compression. On the other hand, \cite{guo_fully_2017} employs a fully passive circuit, achieving higher energy efficiency. Nevertheless, its measurement matrix is limited to a particular type, reducing flexibility for broader applications. Moreover, both designs require one or more ADCs per input channel to support a general measurement matrix, making them unsuitable for photoacoustic and ultrasound imaging RXs, where large-scale sensor arrays necessitate more area- and power-efficient architectures.
In contrast, spatial-domain compression reduces output data rate by allowing multiple channels to share a single ADC, thereby improving area efficiency. Prior work \cite{shoaran_compact_2014}, illustrated in Fig.~\ref{fig:CS_prior_this}~c, applied spatial-domain compression to reduce the data rate in cortical recording applications. However, the design relies on active circuits for MVM, which limits power and area efficiency. To address these limitations, this work adopts spatial domain compression as shown in Fig.~\ref{fig:CS_prior_this}~d. This work presents MVM SAR ADC, which performs passive MVM operations, achieving high energy and area efficiency. The detailed circuit implementation is provided in Section III.

\begin{figure}[t]
\centering
{\includegraphics[width=\linewidth]{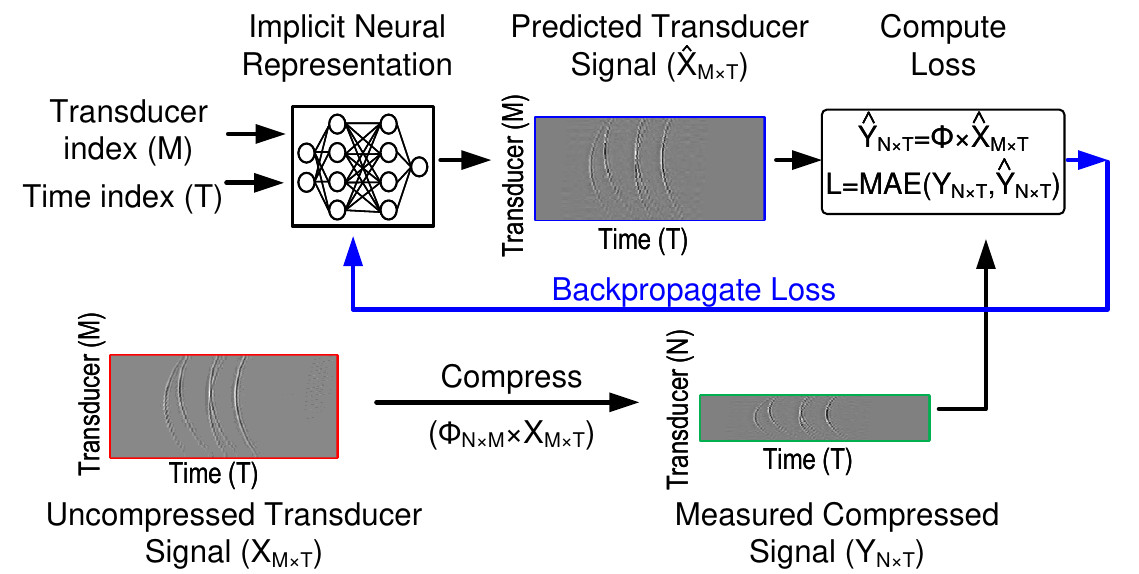}%
} 
\vskip -1ex
\caption{Principles of the INR-based signal reconstruction framework.}
\label{fig:INR_model}
\end{figure}

\begin{figure*}[t!]
\centering
\includegraphics[width=\linewidth]{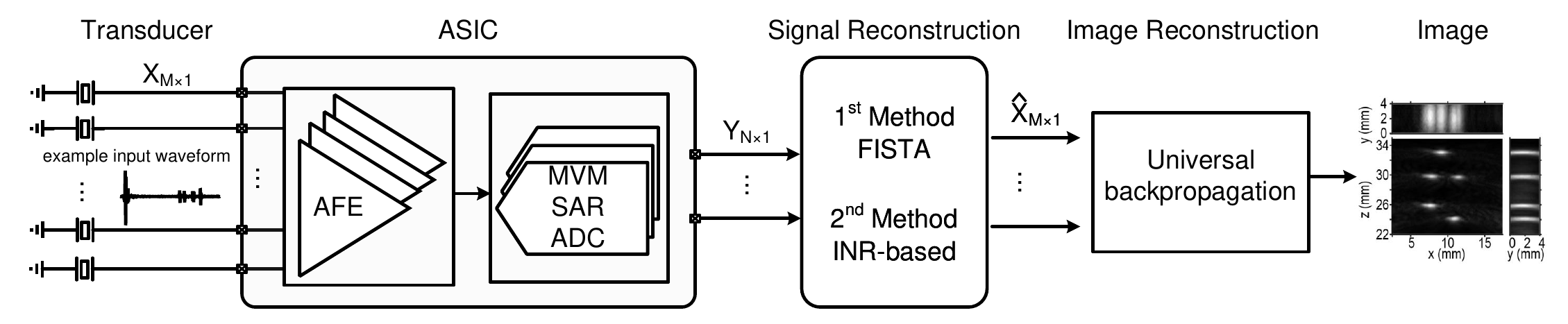}
\vskip -2ex
\caption{Compressive sensing photoacoustic receiver imaging system diagram.}
\label{fig:system_overview}
\end{figure*}

\subsubsection{Measurement Matrix and Reconstruction}
We studied two approaches for designing the measurement matrix and performing signal reconstructions. The first leverages the fast iterative shrinkage-thresholding algorithm (FISTA) \cite{beck_fast_2009} to recover full-array signals from compressed measurements. This method begins with a calibration phase in which uncompressed signals are acquired from the imaging subject. Principal component analysis (PCA) is performed on the uncompressed dataset, and the eigenvectors corresponding to the largest eigenvalues are retained to form the measurement matrix. This process ensures the measurement matrix captures the dominant signal subspace, improving reconstruction accuracy during normal operation. Once the measurement matrix is established, the system transitions to compressed sensing mode, where the hardware multiplies the full-array signal vector \( X \in \mathbb{R}^{M} \) with the reduced-rank measurement matrix \( \Phi \in \mathbb{R}^{N \times M} \) to produce the compressed measurements \( Y \in \mathbb{R}^{N} \). The compressed signals are first transformed into the wavelet domain to enhance sparsity. FISTA is then applied to solve the inverse problem and recover the full-length signal vector \( \hat{X} \), by minimizing the \(\ell_1\)-regularized least squares cost function:

\begin{equation}
\hat{X} = \arg \min_X \frac{1}{2} \|Y - \Phi X\|_2^2 + \lambda \|WX\|_1
\end{equation}

where \( W \) is the wavelet transform matrix and \( \lambda \) is the regularization parameter controlling the tradeoff between data fidelity and sparsity. PCA-optimized measurement matrices and wavelet-domain sparsity enable high-fidelity signal recovery, making this method suitable for applications where the imaging object is known a priori and mostly stationary.

Our second approach adopts the recently proposed implicit neural representation (INR) \cite{sitzmann_implicit_2020,farrell_coir_2023} to reconstruct photoacoustic signals from compressed measurements. INR is a technique that uses a multi-layer perceptron (MLP) neural network to represent signals as continuous functions by implicitly capturing the structure in the signals. Unlike the FISTA-based method, this approach does not require prior acquisition of uncompressed data and instead utilizes a random measurement matrix for general target classes. Reconstructing with INR does not need training datasets and can be optimized with access only to compressed measurements. In our method, as illustrated in Fig.~\ref{fig:INR_model}, we model the full-resolution (uncompressed) transducer signal as an INR, with an MLP architecture of Flexible spectral-bias tuning in Implicit NEural representation (FINER) \cite{liu_finer_nodate}, that takes the transducer index and time index as inputs and outputs the predicted signal value at those coordinates. The full-resolution transducer signal is obtained by evaluating the INR at full-resolution transducer indices and all time indices. The weights of the INR are optimized with compressed measurement as follows. Starting with FINER’s default initialization, at each iteration, the predicted full-resolution transducer signal is generated, then compressed with the chip’s measurement matrix and compared with the compressed measurement. The loss calculated between the predicted compression and the measurement is used to update the weights of the INR using backpropagation and gradient descent. We also use wavelet sparsity to regularize the output of the INR and stabilize the reconstruction. Once the weights of the INR are optimized, the final reconstructed full-resolution uncompressed transducer signal is generated. This method supports flexible adaptation across imaging targets.


\section{System and Circuit Implementation}

\begin{figure}[t]
\centering
{\includegraphics[width=.9\linewidth]{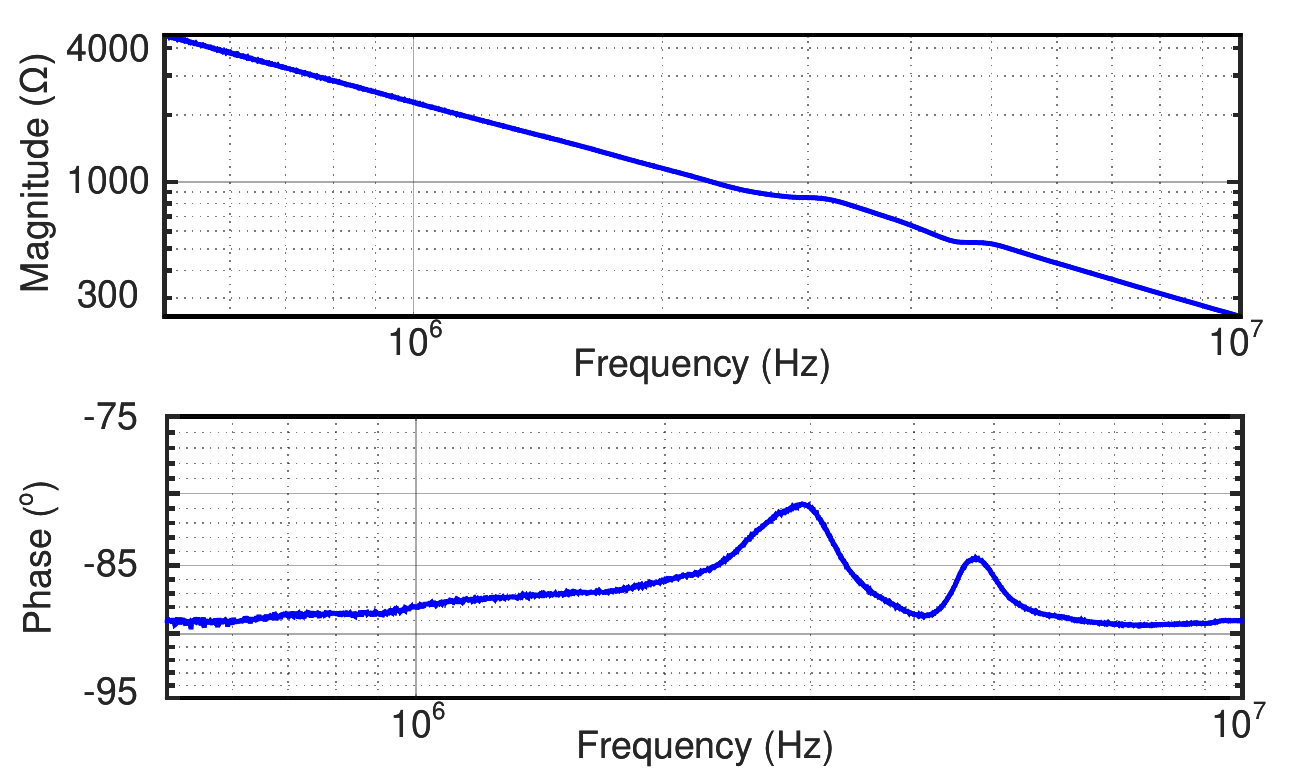}%
} 
\vskip -2ex
\caption{Measured PZT transducer input impedance.}
\label{fig:piezo_impedance}
\end{figure}

\subsection{Imaging Receiver System Overview}
The system comprises three main blocks as shown in Fig.~\ref{fig:system_overview}. The first is the input transducer array, which captures the ultrasound signals. In this prototype, the transducers have a center frequency of 3.5 MHz, fractional bandwidth of 100$\%$, and a pitch of 1 mm. Photoacoustic imaging employs pulsed laser excitation, resulting in the generation of acoustic waves with a corresponding pulsed waveform. To accurately capture the temporal characteristics of these signals, a low quality factor transducer with a broad bandwidth is selected as the receiver. The pitch size is larger than twice the wavelength; this design choice is mainly due to the cost and packaging constraints. The system achieves a depth resolution of 0.5 mm and an in-plane resolution of 1 mm. The input signals from M transducers are represented as a vector \( X\in \mathbb{R}^{M} \).
The measured input impedance of the transducer, shown in Fig.~\ref{fig:piezo_impedance}, is approximately a few hundred ohms near the center frequency.

The second block is the receiver ASIC, which performs compressive sensing by multiplying the input signal vector X with a user-defined measurement matrix (\(\Phi \in \mathbb{R}^{M} \)). The chip carries out the compression as
$Y = Q(\Phi X)$, where \( Y\in \mathbb{R}^{N} \) is the digitized output vector and Q($\cdot$) denotes quantization by the MVM SAR ADC. This reduces the output data rate by a factor of M/N. By integrating compressive sensing directly into the hardware using the MVM SAR ADC, the design achieves enhanced power and area efficiency for the compression operation. The final block is the backend process, where the image is reconstructed using the compressed data. To get the final image, the full-array signal is first reconstructed. Two different methods, discussed in Section II, are used to get the reconstructed full array signal, (\(\hat X \in \mathbb{R}^{M} \)). Once \(\hat X\) is obtained, universal backpropagation is applied to generate the final image.
The backend process is performed off-chip, where high computational resources are available. This allows the system to leverage on-chip compression benefits without imposing significant power constraints on the reconstruction.

\begin{figure*}[!t]
\centering
\includegraphics[width=\linewidth]{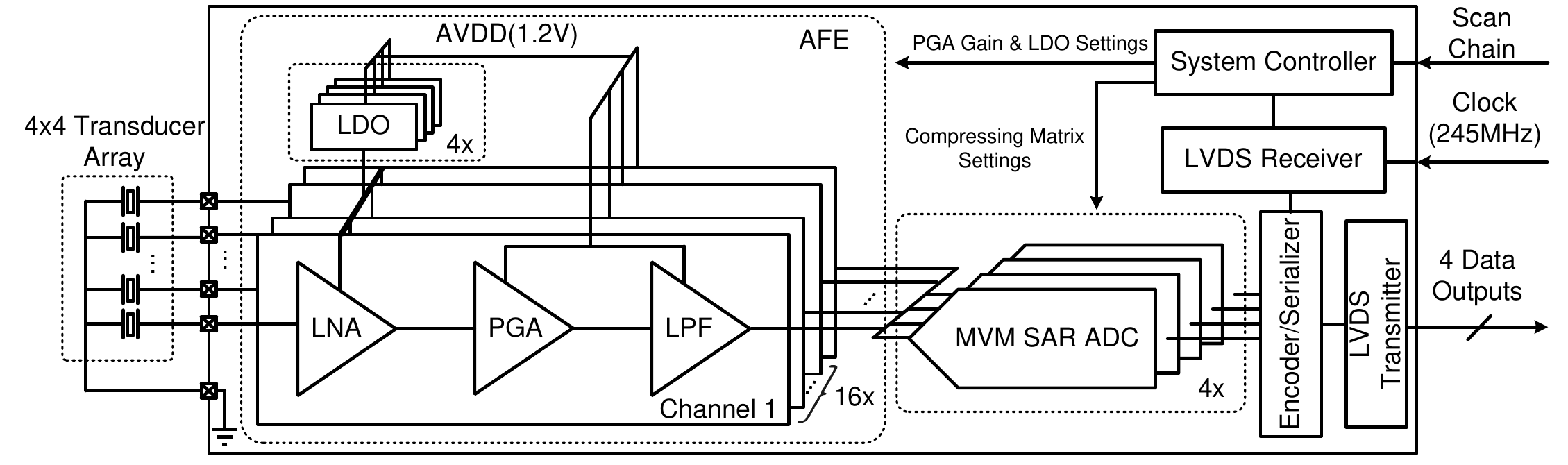}
\vskip -1ex
\caption{System diagram of the compressive sensing PA receiver (RX) chip.}
\label{fig:ASIC_system_overview}
\end{figure*}

\begin{figure}[t]
\centering
{\includegraphics[width=\linewidth]{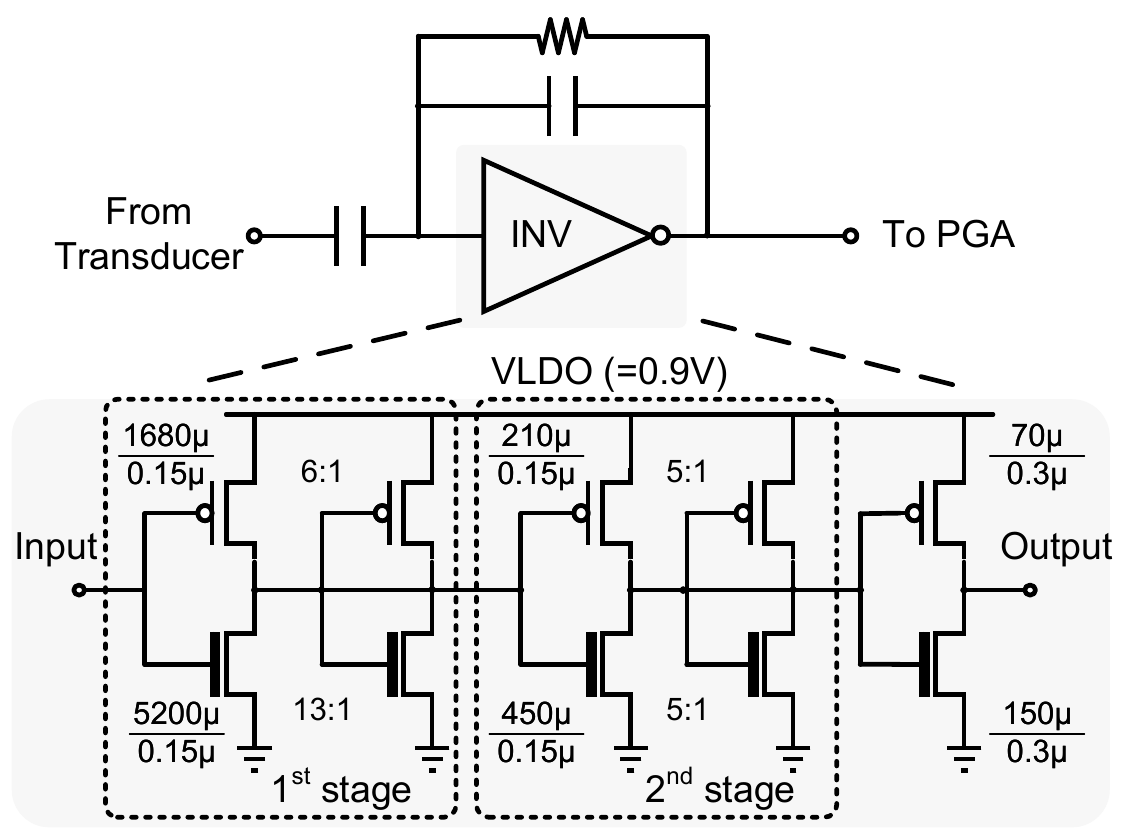}%
} 
\vskip -2ex
\caption{Schematic of LNA.}
\label{fig:LNA}
\end{figure}

\begin{figure}[t]
\centering
{\includegraphics[width=\linewidth]{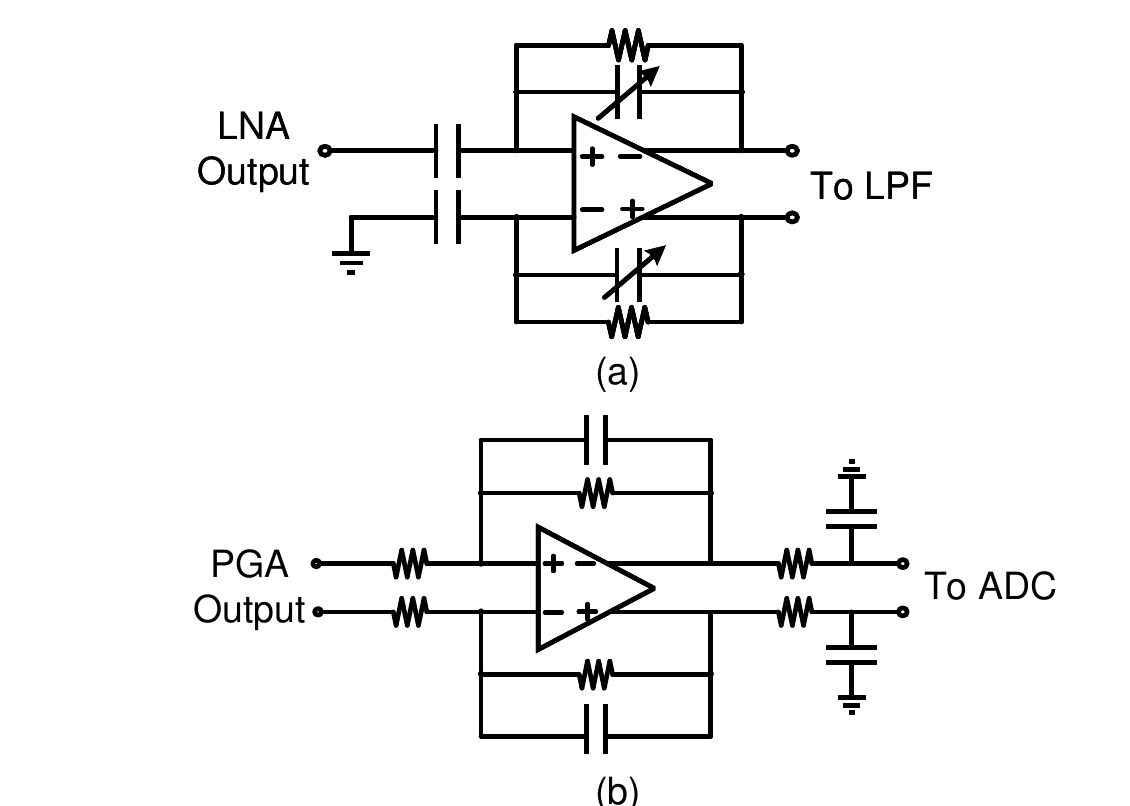}%
}
\vskip -1ex
\caption{Schematics of (a) PGA and (b) filter.}
\label{fig:PGA_FILTER}
\end{figure}

\begin{figure*}[!t]
\centering
\includegraphics[width=.95\linewidth]{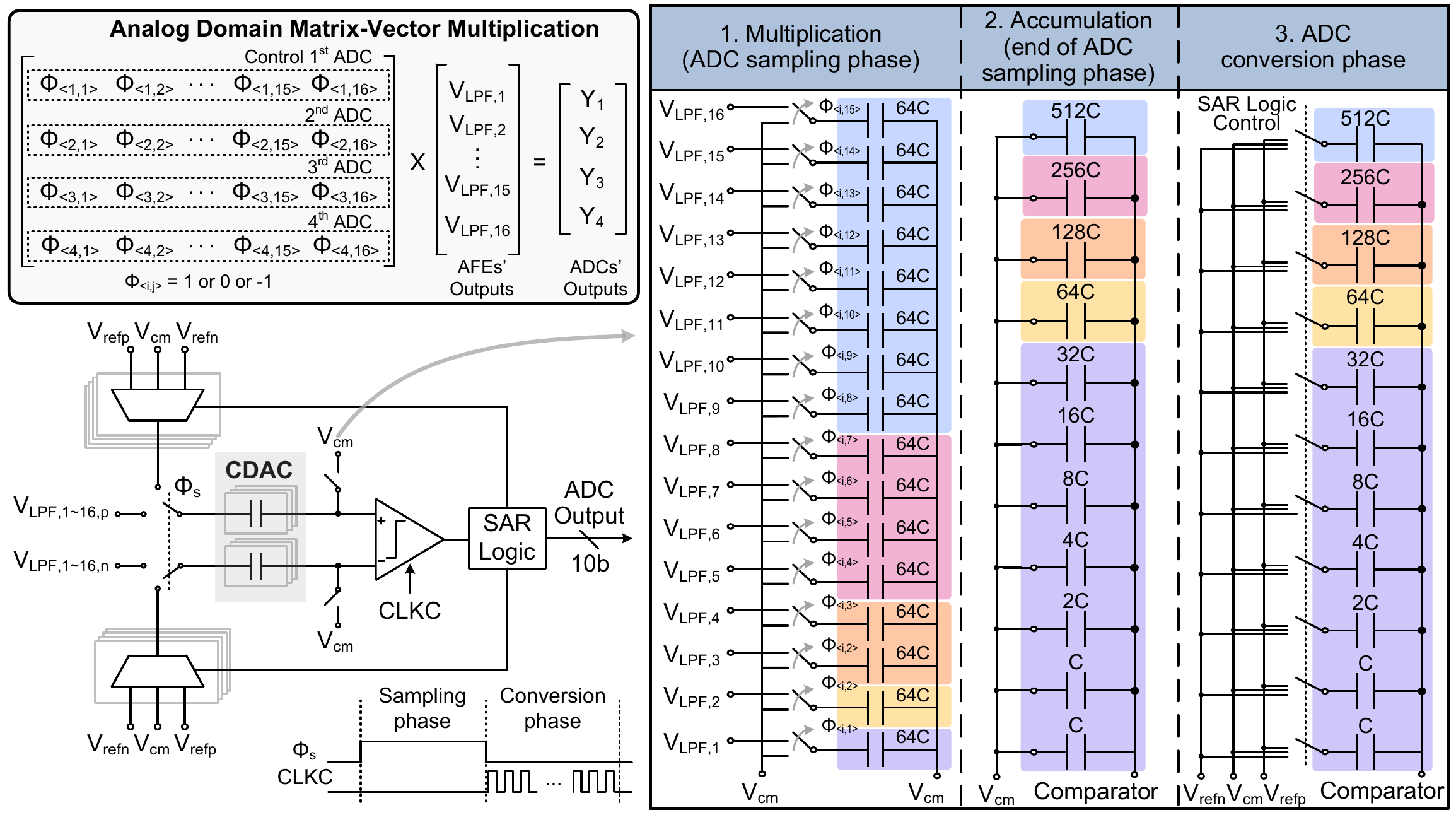}
\vskip -2ex
\caption{Schematic of MVM SAR ADC.}
\label{fig:adc}
\end{figure*}

\subsection{RX circuit design}

Figure~\ref{fig:ASIC_system_overview} illustrates the schematic of the RX. As a proof of concept, a 4-by-4 unfocused transducer array is used as the input (M=16), and the number of output channels is configurable from 1 to 4 (N=1 to 4). The receiver chain consists of a low-noise amplifier (LNA), a programmable gain amplifier (PGA), a low-pass filter (LPF), and the MVM SAR ADC, which performs the compressive sensing operation. 
In typical ultrasound imaging, an SNR of approximately 40 dB is typically sufficient to achieve good image quality \cite{hopf_pitch-matched_2023}. In this work, due to the multi-channel summation in the MVM SAR ADC (detailed in the following section), the noise of each ADC is designed to be 16 times lower than the level corresponding to a 40 dB SNR to preserve single-channel signal quality. This requirement translates to an ADC’s SNDR of 64dB, which is around 10 bits.

\subsubsection{AFE}
Since the input impedance of the transducer is only a few hundred ohms at the center frequency, a voltage amplifier is selected over a transimpedance amplifier for better energy efficiency. The LNA is implemented as a closed-loop, capacitively coupled voltage amplifier to ensure low input-referred noise and consistent channel-to-channel matching, as shown in Fig.~\ref{fig:LNA}. The core of the LNA employs a three-stage inverter-based topology with 1/gm loading, providing robust gain and stable operation without requiring additional bias circuits, similar to the design in \cite{cao_103_2023}. High-threshold voltage NMOS devices are used in the amplifier to suppress flicker noise. Additional flicker noise reduction is achieved by upsizing the NMOS transistor in the first inverter stage, which dominates the overall noise contribution. To maintain the same total area, the NMOS device in the load branch is correspondingly downsized. While this sizing strategy improves noise performance, it introduces a tradeoff: the smaller load device limits the maximum current flowing, which can degrade the amplifier’s transient response when subjected to large signal steps, particularly from high to low voltage levels.
However, in the context of photoacoustic imaging, the input signal is a band-limited signal with no abrupt high-to-low voltage transitions. As such, the LNA topology is well-suited to the application. The first stage of the amplifier has a nominal transconductance of 56.8 mS with a 1.2 mA bias current. To improve power supply rejection ratio (PSRR), four LNAs share a common low-dropout regulator (LDO). The LDO has off-chip decoupling capacitors to ensure stable operation and enhanced PSRR across the target frequency band.

Acoustic waves typically experience about 1 dB/cm/MHz attenuation in soft tissue \cite{hoskins2019diagnostic}, while light suffers greater attenuation and scattering, limiting imaging depth to a few centimeters\cite{ermilov2009laser}. To compensate for depth-dependent signal loss, this work uses a PGA with a gain range of 18 dB for approximately 2 cm depth. In future designs, the PGA can be improved by adopting time-gain control \cite{guo_125_2024}. The PGA employs a two-stage feedforward operational amplifier \cite{wang_reconfigurable_2018} to enhance energy efficiency and supports selectable gain settings of 8, 16, 32, or 64 V/V. It also converts the single-ended output of the LNA into a differential signal, improving common-mode rejection and expanding the dynamic range for subsequent stages. Following the PGA, a first-order RC LPF is implemented using a two-stage Miller-compensated operational amplifier. The LPF provides an additional 2 times voltage gain and acts as an anti-aliasing filter for the following ADCs. The 3dB bandwidth of the LPF is 10.35MHz.

\subsubsection{MVM SAR ADC}
The MVM SAR ADC schematic is shown in Fig.~\ref{fig:adc}. The required MVM operation is multiplying a 4-by-16 measurement matrix with the 16-channel input to produce a compressed 4-by-1 output. The measurement matrix supports ternary weight settings ($-1$,0,1), enabling flexible and reconfigurable sensing. The operation is carried out using four parallel MVM SAR ADCs, each responsible for computing the dot product between one row of the measurement matrix and the input vector. 
The ADC is designed with asynchronous logic and a bottom-plate sampling scheme, producing a 10-bit final output. The asynchronous design eliminates the need for high-speed global clock distribution, thereby reducing power consumption and design complexity. The comparator employs with a dynamic-bias preamplifier using tail charge pump \cite{bindra_174vrms_2022} to achieve high speed, low noise, and high energy efficiency. Multiplication and accumulation are carried out passively using the sampling capacitor within the SAR ADC. During the sampling phase, the sampling capacitor is segmented into 16 equally weighted capacitors. Each capacitor samples one of three inputs based on the assigned weight: the positive or negative differential output of the corresponding LPF (for +1 or -1), or the common-mode voltage, $V_{cm}$ (for 0).

The charge store on each 16 equally weighted capacitor bank ($Q_i$) is:
\begin{equation}
Q_i=\Phi_i\times (V_{LPF,i}-V_{cm})\times 64C\quad i=1\sim16
\end{equation}
where C, $\Phi$ and $V_{LPF}$ are the unit capacitance, measurement matrix, and output of AFEs, respectively.
This configuration effectively implements the analog-domain multiplication. At the end of the sampling phase, the top-plate sampling switch is disconnected, and the bottom plates of all capacitors are connected to $V_{cm}$. The charge is redistributed among the capacitors, equivalent to accumulating the multiplication results. The multiplication and accumulation (MAC) result becomes the top-plate voltage as 
\begin{equation}
V_{top} = V_{cm}- \sum_{i=1}^{16}Q_i\times \frac{1}{1024C}=2V_{cm}-\sum_{i=1}^{16}\Phi_i\times V_{LPF,i}
\end{equation}
Then, the sampling capacitor is split into binary weights. Finally, the standard SAR conversion process, switching the bottom plate voltage to positive or negative reference voltage ($V_{refp/n}$) based on comparator results, produces the final result. The MVM SAR ADC achieves a fully passive, area- and power-efficient MAC operation that supports arbitrary ternary-weighted matrix multiplication. The sampling capacitor is 0.8 pF per side, designed to meet the thermal noise requirement. Following the mismatch calculation in \cite{harpe_26_2011}, the sampling capacitor mismatch is around 10 times smaller than the specified linearity limit, ensuring negligible impact on overall performance.

\section{Experimental Results}
\begin{figure}[t]
\centering
{\includegraphics[width=\linewidth]{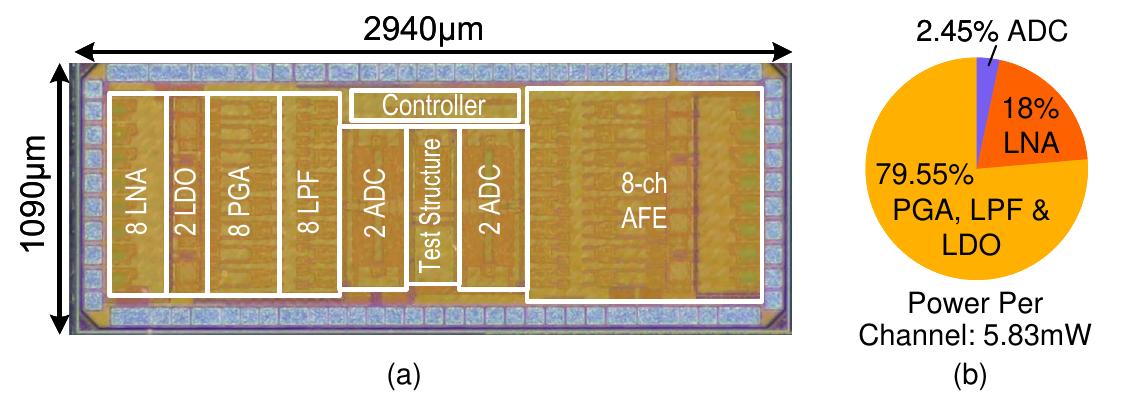}%
} 
\vskip -2ex
\caption{(a) Micrograph and (b) power breakdown of the test chip.}
\label{fig:die_photo}
\end{figure}

The chip is fabricated in TSMC 65nm CMOS
technology, occupying an area of 0.118 $mm^2$ per channel (see Fig.~\ref{fig:die_photo}~a). Figure \ref{fig:die_photo}~b shows the power breakdown for one channel. The performance of the chip is first measured, followed by phantom imaging to evaluate system-level functionality.

\subsection{Chip Testing}

\begin{figure}[t]
\centering
{\includegraphics[width=0.85\linewidth]{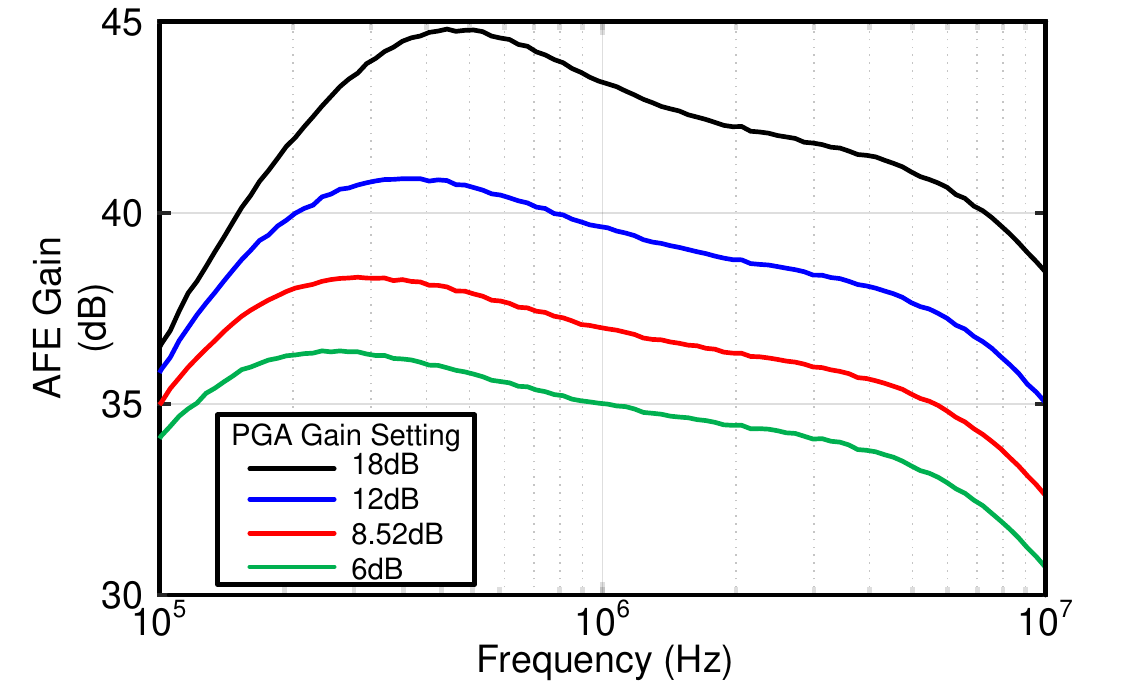}%
}
\vskip -2ex
\caption{Measured AFE AC response.}
\label{fig:AFE_AC_response}
\end{figure}

\begin{figure}[t]
\centering
\vskip -1ex
{\includegraphics[width=0.85\linewidth]{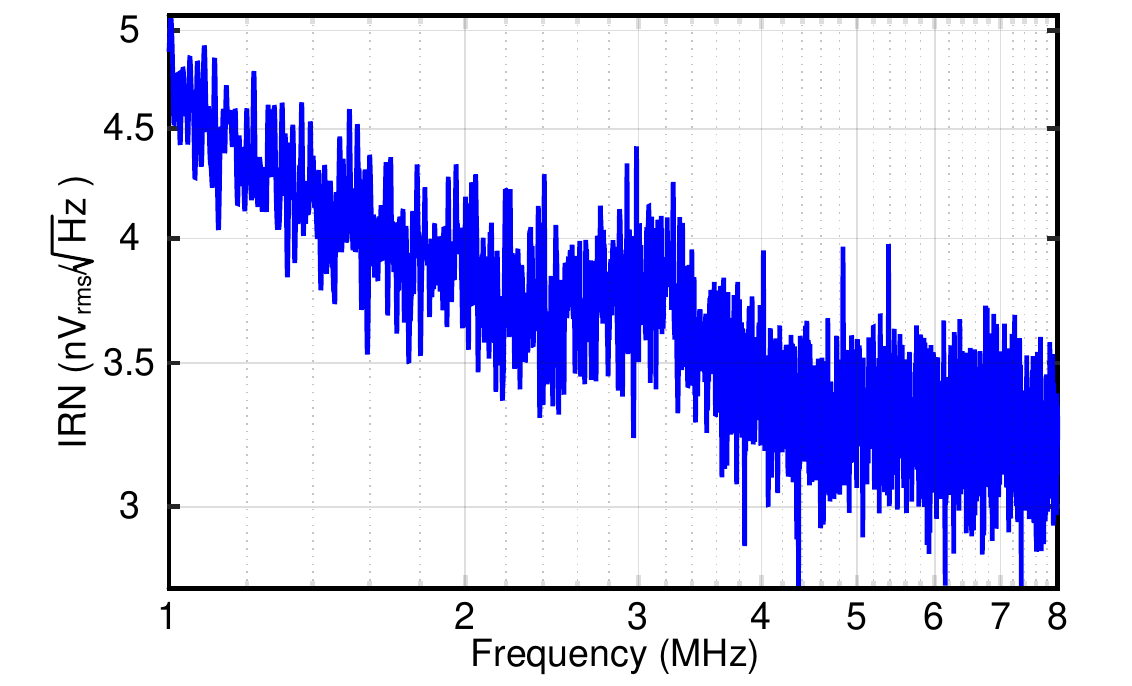}%
}
\vskip -2ex
\caption{Measured AFE noise spectrum.}
\label{fig:AFE_noise}
\end{figure}

\begin{figure}[t!]
\centering
\vskip -1ex
{\includegraphics[width=0.85\linewidth]{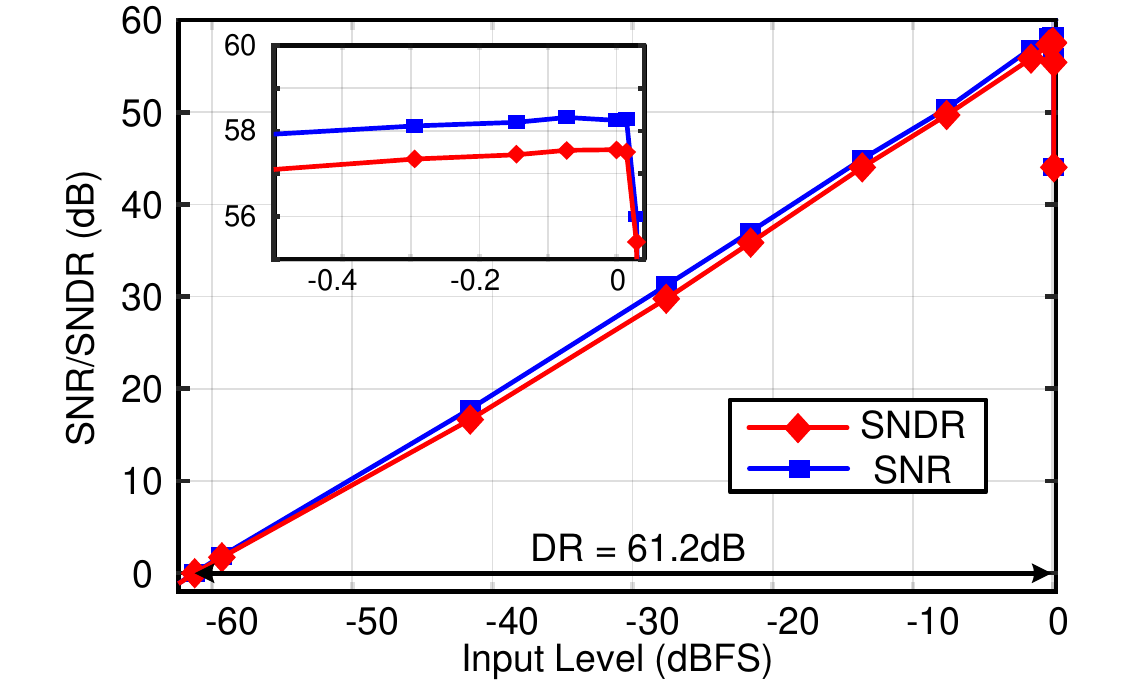}%
} 
\vskip -2ex
\caption{Measured SNDR/SNR versus input signal magnitude.}
\label{fig:ADC_SNR_SNDR}
\end{figure}

\begin{figure}[t]
\centering
{\includegraphics[width=.85\linewidth]{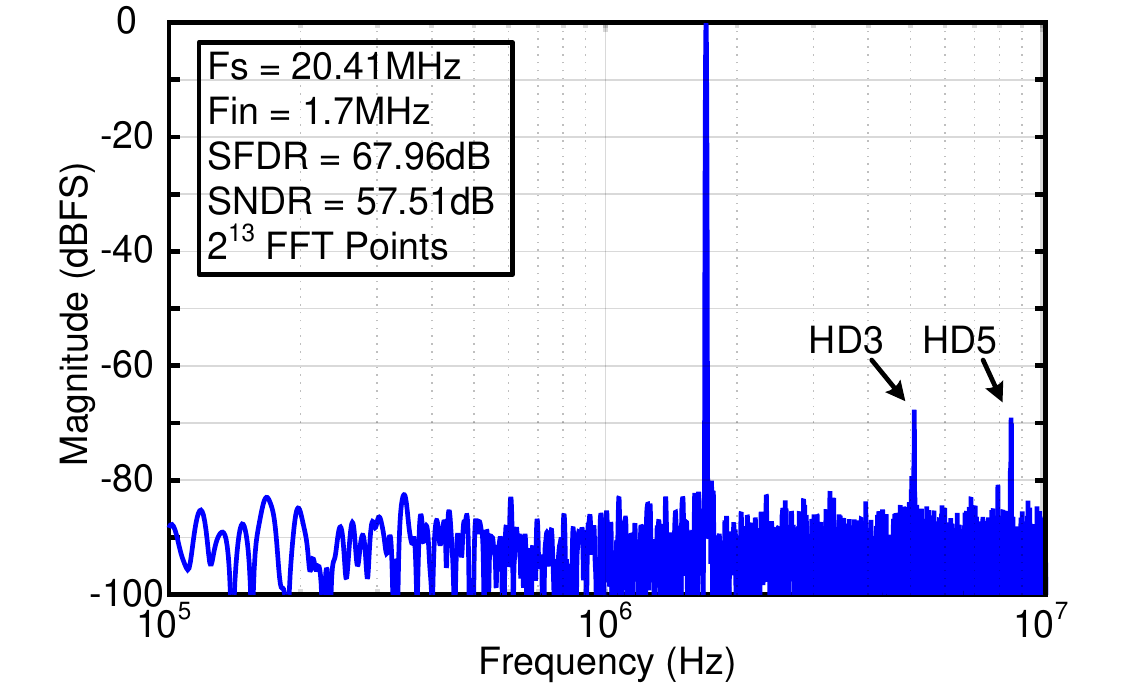}%
}
\vskip -2ex
\caption{Measured spectrum of the ADC with 1.7-MHz input signal.}
\label{fig:ADC_spectrum}
\end{figure}

The AFE’s AC response is shown in Fig.~\ref{fig:AFE_AC_response}. At the highest gain setting, the gain at the center frequency reaches 41.7 dB. The gain flatness within the target frequency band (1.75 MHz to 5.25 MHz) is approximately 1 dB. A gain peaking is observed near 400 kHz, likely due to complex poles introduced by parasitic capacitance, degrading the phase margin of the amplifier. This peaking is effectively suppressed by the transducer's bandpass characteristic and a digital bandpass filter applied in post-processing, resulting in negligible impact on image quality. While this issue can be addressed in future designs, it does not affect the performance of the current imaging system. The measured input-referred noise floor of the AFE, shown in Fig.~\ref{fig:AFE_noise}, is approximately 3.5nV/$\sqrt{Hz}$. The elevated noise at lower frequencies is primarily attributed to the flicker noise of the NMOS transistor in the first stage of the LNA. This flicker noise corner can be shifted to lower frequencies by further upsizing the transistor. However, increasing the transistor size also raises its parasitic capacitance, Advanced techniques such as chopping or auto-zeroing can be considered in future design.
Fig.~\ref{fig:ADC_SNR_SNDR} presents the measured SNR and SNDR versus input amplitude, and the measured dynamic range of 61.2 dB. Operating at a 20.41 MHz sampling rate, the ADC achieved an SNDR of 57.51dB without calibration, corresponding to an ENOB of 9.26. The measured output spectrum is shown in Fig.~\ref{fig:ADC_spectrum}.

\begin{figure}[t]
\centering
\vskip -1ex
{\includegraphics[width=.85\linewidth]{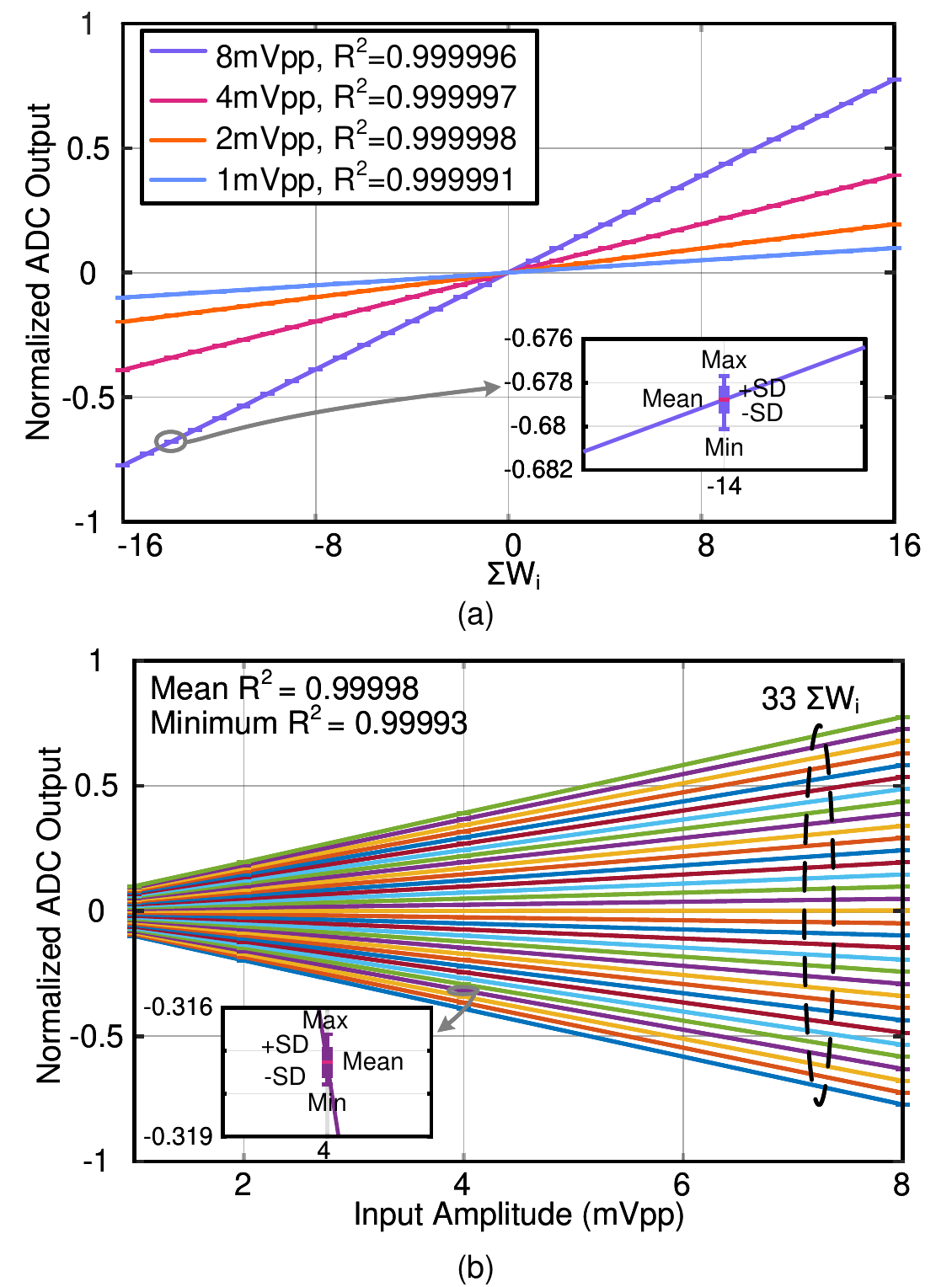}%
}
\vskip -2ex
\caption{(a) Fixed Input and Sweep $\Sigma$Wi; (b) Fixed $\Sigma$Wi and Sweep Input.}
\label{fig:computing_linearity}
\end{figure}

Next, we evaluated the computing linearity of the compressive sensing RX, which is affected by the nonlinearity introduced by the AFE, ADC, and mismatches among channels and capacitors. The MVM operation can be represented as $Output = \Sigma(W_i\times I_j)$, where $W_i$ and $I_j$ denote the weight and the input for each channel, respectively. To evaluate the computing linearity, the same 3.5MHz sine wave input is applied across all channels. The output can then be expressed as
$Output = \Sigma(W_i)\times I$. Computing linearity is evaluated by sweeping the weight sum ($\Sigma W_i$) settings and input ($I$) amplitude. First, $\Sigma W_i$ is swept from its minimum to maximum values, spanning from -16 to 16 (e.g., 16 corresponds to all weights set to 1). For each $\Sigma W_i$ value, up to 50 weight combinations are selected randomly from all possible configurations that yield the same $\Sigma W_i$. This experiment is performed for four AFE input amplitudes (1mVpp, 2mVpp, 4mVpp, 8mVpp), and the resulting output is plotted against $\Sigma W_i$ in Fig.~\ref{fig:computing_linearity}~a. The results demonstrate high MVM linearity with respect to weight, with a minimum coefficient of determination ($R^2$) of 0.999991 across four input amplitudes. Next, $\Sigma W_i$ is fixed while sweeping the input to evaluate linearity with respect to the input, $I$. The corresponding results are shown in Fig.~\ref{fig:computing_linearity}~b. The minimum $R^2$ is 0.99993 across all 33 possible $\Sigma W_i$ values. The case $\Sigma W_i=0$ is excluded for $R^2$ calculation, since the output is constant, rendering $R^2$ undefined.

\subsection{Imaging Testing}
\begin{figure}[t]
\centering
{\includegraphics[width=\linewidth]{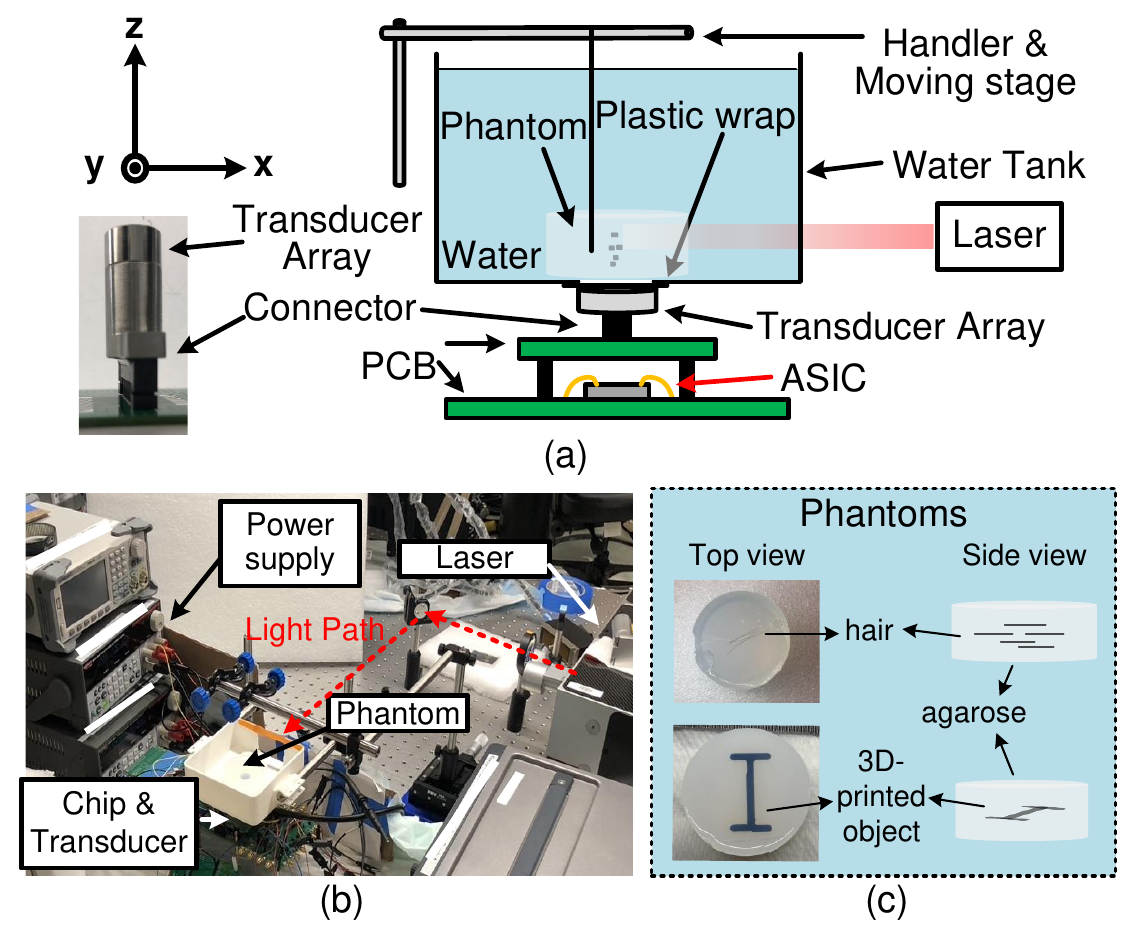}%
}
\vskip -2ex
\caption{Photoacoustic imaging setup (a) diagram, (b) picture, (c) phantoms. }
\label{fig:Setup_phantom}
\end{figure}

To validate the functionality of the compressive sensing RX, photoacoustic imaging experiments were conducted using two custom-designed phantoms. The experimental setup is shown in Fig.~\ref{fig:Setup_phantom}~(a,b). The phantoms were submerged in a water tank and held in place using a 3D-printed holder, with the transducer array and receiver chip positioned at the bottom of the tank. A pulsed laser source was directed from the side to illuminate the phantom. The two phantom photos are shown in Fig.~\ref{fig:Setup_phantom}~c. The first phantom consists of five human hairs embedded in an agarose gel to mimic blood vessel structures. A 24-by-4 array configuration is emulated by moving the phantom along the y-axis in six discrete steps, with the signal acquisition at each position performed independently. This approach allows the system to simulate a larger array using a single chip and demonstrates that the effective measurement matrix in this case becomes block-diagonal. Such a configuration could be directly mapped to a multi-chip architecture in future designs, enabling simultaneous operation for larger-scale imaging. The excitation source is a pulsed laser with a 750 nm wavelength and 6 ns pulse width, operating at 20 Hz, and delivering an average fluence of 1.9 mJ/$cm^2$ per pulse. To reduce the effect of laser energy variation between acquisitions, each acquisition was averaged over 16 laser pulses.

\begin{figure}[t]
\centering
{\includegraphics[width=1\linewidth]{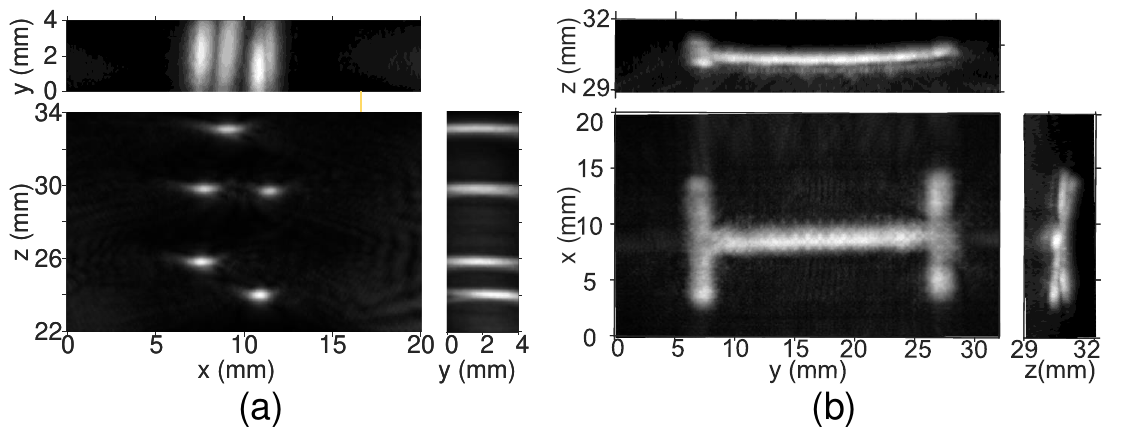}%
}
\vskip -2ex
\caption{Reconstructed images using uncompressed data (a) 5 hairs (b) I-shaped 3D-printed object.}
\label{fig:uncompressed}
\end{figure}

\begin{figure*}[!t]
\centering
\includegraphics[width=\linewidth]{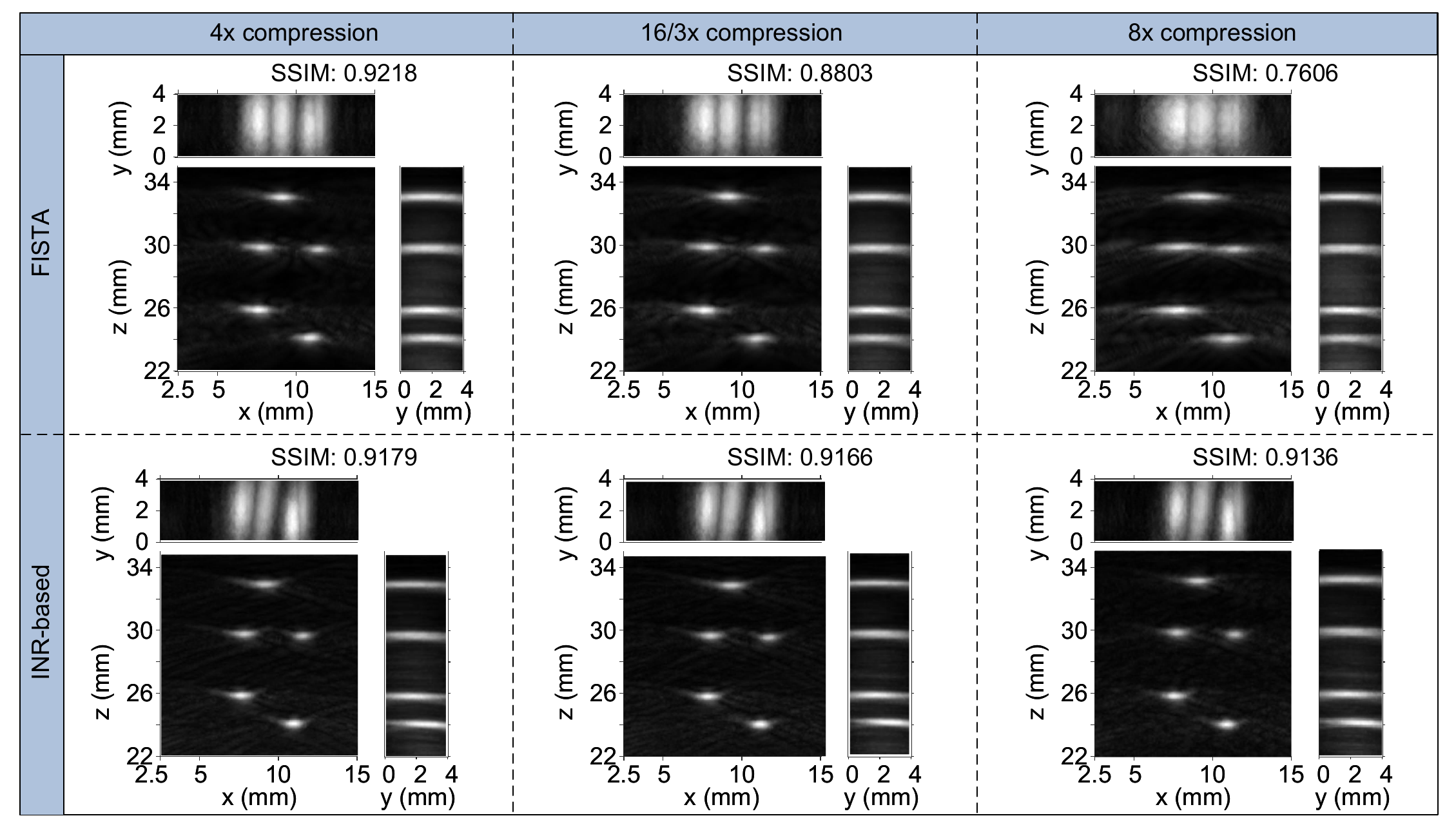}
\vskip -2ex
\caption{5 hair phantom reconstructed images.}
\label{fig:5_hair_image_compressed}
\end{figure*}

\begin{figure*}[!t]
\centering
\includegraphics[width=\linewidth]{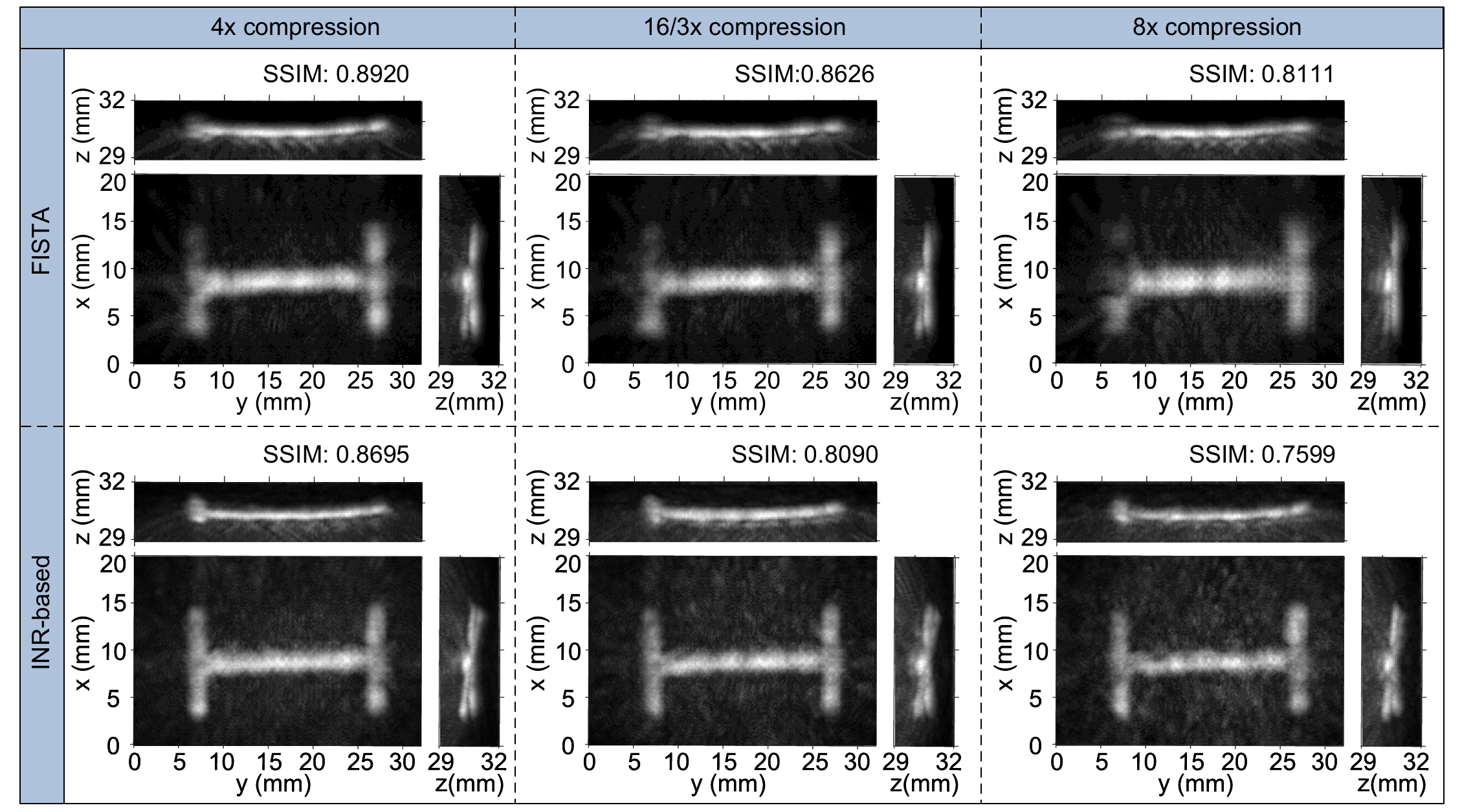}
\vskip -2ex
\caption{I-shape reconstructed images.}
\label{fig:I_shape_compressed}
\end{figure*}

\begin{figure}[t]
\centering
{\includegraphics[width=1\linewidth]{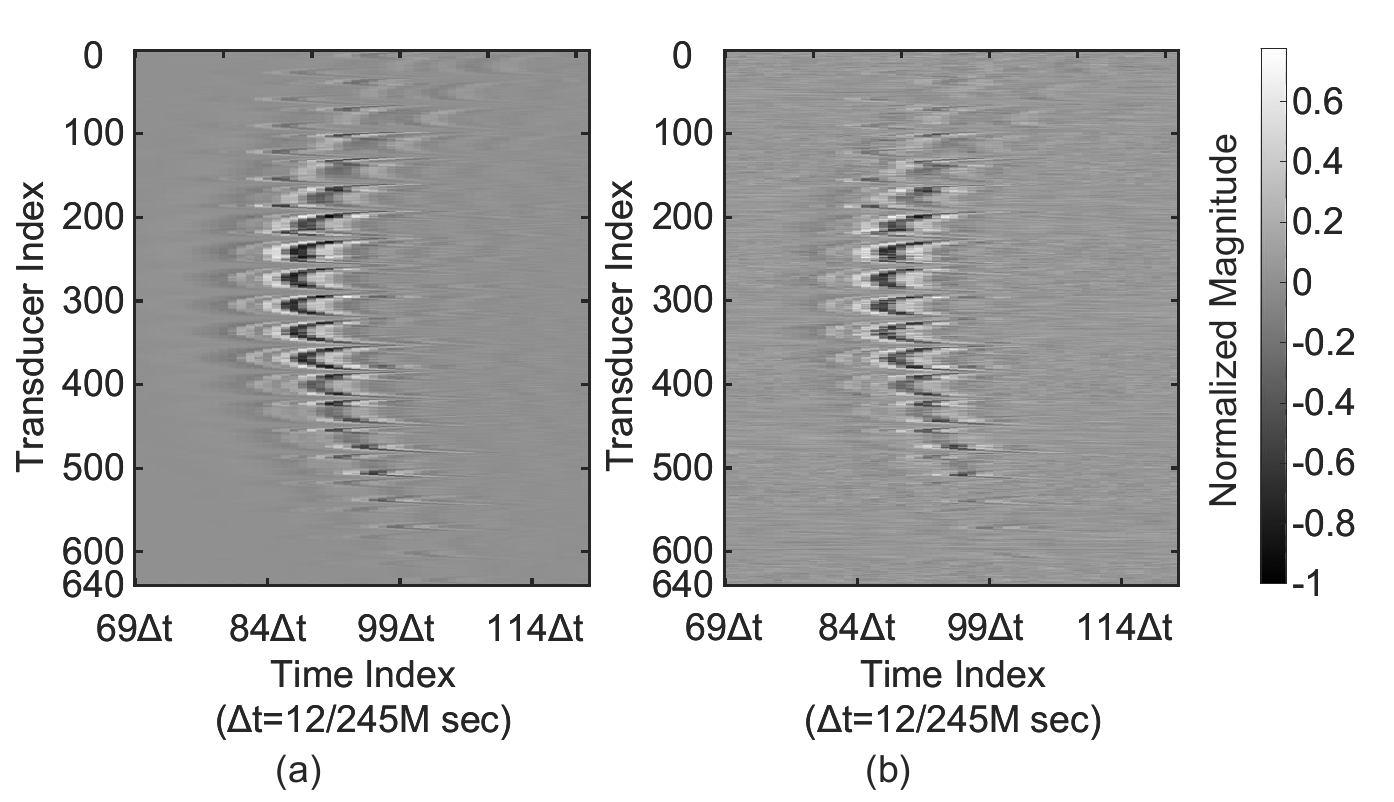}%
} \\
\hfil
\vskip -2ex
\caption{Signal comparison between (a) raw data (b) 4x compressive sensing.}
\label{fig:signal_comparison}
\end{figure}

The second phantom features a 3D-printed I-shaped object designed to evaluate performance on higher-contrast, structured targets. A 24-by-32 array configuration is emulated by scanning the phantom across an 8-by-4 grid in both x- and y-directions. During scanning, the 4-by-4 array is mechanically shifted by 4 mm in either the x- or y-direction in both phantom experiments, effectively emulating a larger array with a transducer pitch of 1 mm. The excitation laser delivers an average fluence of 19 $\mu$J/cm² per pulse. The reduced laser power is selected to accommodate the strong photoacoustic response of the 3D-printed material, ensuring that the input signal remains within the dynamic range of the RX system. Uncompressed data is obtained by configuring the measurement matrix with a single '1' in each row, and cycling its position across all 16 channels to acquire the full data. The reconstructed three-view images from the uncompressed data for both phantoms are shown in Fig.~\ref{fig:uncompressed}. All images are shown directly from reconstruction without applying any further post-processing. Clear structural features corresponding to the five-hair phantom and the I-shaped object are distinctly visible.

Next, reconstructed phantom images using compressed data are shown in Fig.~\ref{fig:5_hair_image_compressed} and Fig.~\ref{fig:I_shape_compressed} for three different compression ratios. Two reconstruction methods, FISTA-based and INR-based, were evaluated. We updated the original INR results for the 5 hairs phantom in \cite{liao_352_2025} with images reconstructed using the updated INR model. Both approaches demonstrate high-quality image recovery at compression ratios of 4x and 16/3x, successfully preserving key structural features of the phantom. At the highest tested compression ratio of 8x, the INR-based method exhibits superior spatial localization and reduced signal spread compared to FISTA, highlighting its robustness under more aggressive compression. The 3D structural similarity index (SSIM) between the reconstructed images using compressed and uncompressed data are provided in Fig.~\ref{fig:5_hair_image_compressed} and Fig.~\ref{fig:I_shape_compressed}. These results confirm the effectiveness of the system in maintaining image fidelity across a range of compression levels.
Figure~\ref{fig:signal_comparison} shows an example of the raw signal with the signal reconstructed using INR at a 4x compression ratio. The reconstructed signal closely preserves the key features of the original waveform, including the overall structure and temporal coherence, although some high-frequency components are slightly degraded due to reconstruction noise.

\begin{figure}[t]
\centering
{\includegraphics[width=1\linewidth]{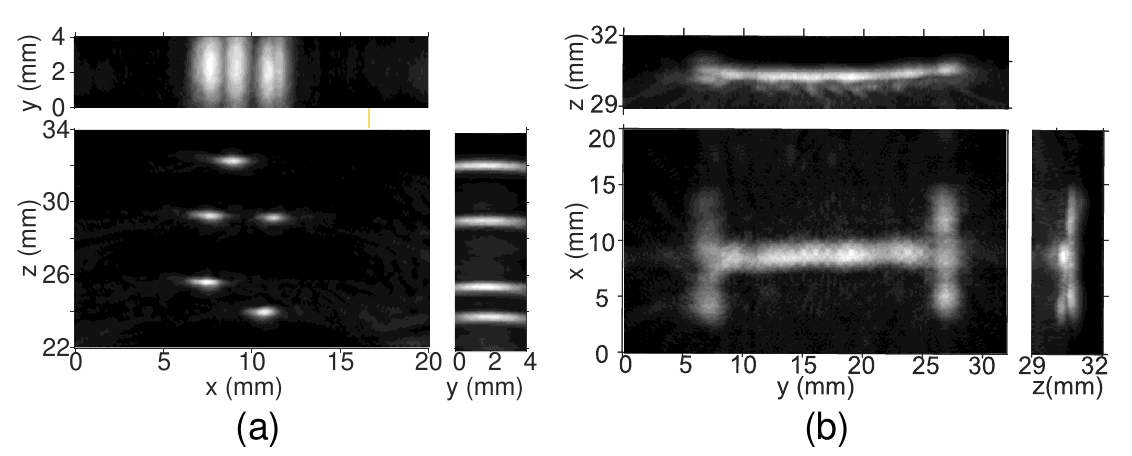}%
} \\
\hfil
\vskip -2ex
\caption{Reconstructed images using software-emulated compressed data (a) 5 hairs (b) I-shaped 3D-printed object.}
\label{fig:software_emulated}
\end{figure}

To evaluate the fidelity of the on-chip compression, the entire compressive sensing process was emulated in software using the same measurement matrices applied during hardware acquisition. The reconstructed images from the software-emulated compression are shown in Fig.~\ref{fig:software_emulated} and are directly compared with those obtained from hardware measurements. There is minimal difference between the two sets of results, confirming that the receiver architecture performs accurate analog-domain compression with negligible degradation. Quantitatively, the SSIM between software-emulated and hardware-based compression is 0.9426 for the five-hair phantom and 0.9696 for the I-shaped phantom. This validation underscores the reliability of the MVM SAR ADC and the overall system in preserving image quality under compressed acquisition.

To evaluate the hardware cost for reconstruction, the INR neural network has around 200k trainable parameters and is iteratively trained for 400 iterations. For full sensor resolution of 640 transducers used in the I-shaped phantom experiment, the training process takes 2.6 seconds on an RTX 4090 GPU. Signal reconstruction using the FISTA method takes less than 50 milliseconds on an Intel i9 processor.

\subsection{Discussion on Scalability and Wearable Implementation}

Scaling to larger arrays is achievable by replicating RX channels. Our phantom experiment demonstrates that a block-diagonal sensing matrix supports modular expansion. While we emulate a larger array using a single chip moved across positions, future implementations can tile multiple chips or integrate more channels on-chip for higher frame rates. Block size can be tuned to the spatial characteristics of the target. A larger block enables higher compression and relax sensing matrix constraints. However, increasing channels per ADC introduces circuit-level challenges, including greater inter-channel mismatch from smaller CDACs, and higher parasitics from additional switches and routing—leading to increased power and crosstalk. 
Using higher-frequency transducers improves spatial resolution but also increases tissue attenuation, requiring wider bandwidth, lower-noise RX, and advanced transducer packaging \cite{guo_125_2024}.
 
As array size or frequency increases output data rates increase, making compression and efficient processing critical. In this work, FISTA scales linearly with input size assuming fixed iteration count, but its per-iteration cost is dominated by MVM, which can become a bottleneck for large datasets. A single MLP models the uncompressed data, with complexity scaling nonlinearly with data size. For large datasets, model size can be reduced using multiple small MLPs \cite{Reiser_2021_ICCV} or hierarchical representations \cite{saragadam2022miner}.

Integrating a pulsed laser into a wearable form factor poses significant challenges due to its high power consumption and bulky packaging. As a more practical alternative, LED-based light sources—demonstrated in prior works \cite{zhu_light_2018}, \cite{zhu_towards_2020}—offer advantages for wearable applications, including compact size, lower cost, and substantially reduced power requirements. High-power LED arrays are capable of delivering pulse energies up to 200 µJ with pulse widths ranging from tens to hundreds of nanoseconds and repetition rates of several kilohertz. This high repetition rate enables signal averaging to achieve adequate imaging quality. The typical power consumption of such LED arrays during operation is on the order of a few hundred milliwatts. Moreover, the small footprint of individual LEDs ($\sim$1 mm²) facilitates the design of dense array configurations well-suited for integration into compact and flexible wearable systems.

\subsection{Comparison with Prior Arts}

Table \ref{tab1} compares this work to the state of the art. Compared to the state-of-the-art photoacoustic RX in \cite{chen_pixel_2017}, our design achieves nearly 4 times improvement in power efficiency while maintaining a comparable dynamic range. Since the amplitudes of photoacoustic signals are typically weaker than those of ultrasound signals, the receiver demands a lower input-referred noise floor. This work meets that requirement by achieving an input-referred noise of 3.5 nV/$\sqrt Hz$ with a per-channel power consumption of 5.83 mW. Furthermore, the use of analog-domain compressive sensing significantly reduces the output data rate while preserving low-loss full-array information, enabling accurate image reconstruction.

\begin{table*}[t]
\caption{\textbf{Comparisons with State-of-the-Art Ultrasound/Photoacoustic Imaging Receiver}}
\label{table}
\centering
\setlength{\tabcolsep}{2pt}
\renewcommand{\arraystretch}{1.5}
\begin{tabular}{|p{100pt}|p{52pt}|p{50pt}|p{50pt}|p{50pt}|p{50pt}|p{50pt}|p{50pt}|}
\hline

& 
\parbox[c][0.8cm]{50pt}{\centering
{\textbf{This Work}}}&
\parbox[c][0.8cm]{52pt} {\centering{M.-C. Chen \\JSSC'17\cite{chen_pixel_2017}}} &
\parbox[c][0.8cm]{50pt} {\centering{A. Sawaby\\VLSI'18\cite{sawaby2018wireless}}} &
\parbox[c][0.8cm]{50pt} {\centering{J. Li\\VLSI'19\cite{li_154mwelement_2019}}} &
\parbox[c][0.8cm]{50pt} {\centering{Y. Hopf\\ISSCC'22\cite{hopf_pitch-matched_2022}}} &
\parbox[c][0.8cm]{50pt} {\centering{P. Guo\\JSSC'24\cite{guo_125_2024}}} &
\parbox[c][0.8cm]{50pt} {\centering{J. Lee\\JSSC'21\cite{lee_36-channel_2021}}}
\\\hline

\multicolumn{1}{|c|}{\textbf{Technology}} 
& \multicolumn{1}{c|}{\textbf{65nm}} 
& \multicolumn{1}{c|}{28nm} 
& \multicolumn{1}{c|}{65nm}
& \multicolumn{1}{c|}{180nm}
& \multicolumn{1}{c|}{180nm BCD}
& \multicolumn{1}{c|}{180nm BCD}
& \multicolumn{1}{c|}{180nm}
\\\hline

\multicolumn{1}{|c|}{\textbf{Imaging Modality}} 
& \multicolumn{1}{c|}{\textbf{Photoacoustic}} 
& \multicolumn{1}{c|}{Photoacoustic} 
& \multicolumn{1}{c|}{Thermoacoustic}
& \multicolumn{1}{c|}{Ultrasound}
& \multicolumn{1}{c|}{Ultrasound}
& \multicolumn{1}{c|}{Ultrasound}
& \multicolumn{1}{c|}{Ultrasound}
\\\hline

\multicolumn{1}{|c|}{\textbf{Transducer}} 
& \multicolumn{1}{c|}{\textbf{PZT}} 
& \multicolumn{1}{c|}{CMUT} 
& \multicolumn{1}{c|}{CMUT}
& \multicolumn{1}{c|}{PZT}
& \multicolumn{1}{c|}{PZT}
& \multicolumn{1}{c|}{PZT}
& \multicolumn{1}{c|}{PMUT}
\\\hline

\multicolumn{1}{|c|}{\textbf{Transducer Array Size}} 
& \multicolumn{1}{c|}{\textbf{4x4}} 
& \multicolumn{1}{c|}{4x4} 
& \multicolumn{1}{c|}{16x1}
& \multicolumn{1}{c|}{4x4}
& \multicolumn{1}{c|}{8x9}
& \multicolumn{1}{c|}{16x16}
& \multicolumn{1}{c|}{6x6}
\\\hline

\multicolumn{1}{|c|}{\textbf{Center Frequency (MHz)}} 
& \multicolumn{1}{c|}{\textbf{3.5}} 
& \multicolumn{1}{c|}{5} 
& \multicolumn{1}{c|}{10}
& \multicolumn{1}{c|}{5}
& \multicolumn{1}{c|}{6}
& \multicolumn{1}{c|}{9}
& \multicolumn{1}{c|}{5}
\\\hline

\multicolumn{1}{|c|}{\textbf{Nyquist Sampling Rate (MHz)}} 
& \multicolumn{1}{c|}{\textbf{20.41}} 
& \multicolumn{1}{c|}{20} 
& \multicolumn{1}{c|}{30}
& \multicolumn{1}{c|}{30}
& \multicolumn{1}{c|}{24}
& \multicolumn{1}{c|}{40}
& \multicolumn{1}{c|}{20}
\\\hline

\multicolumn{1}{|c|}{\textbf{Input-Referred Noise}} 
& \multicolumn{1}{c|}{\textbf{3.5 nV/$\mathbf{\sqrt{Hz}}$}} 
& \multicolumn{1}{c|}{N/A} 
& \multicolumn{1}{c|}{4 mPa/$\sqrt{Hz}$}
& \multicolumn{1}{c|}{N/A}
& \multicolumn{1}{c|}{N/A}
& \multicolumn{1}{c|}{0.7 pA/$\sqrt{Hz}$}
& \multicolumn{1}{c|}{19.3 nV/$\sqrt{Hz}$}
\\\hline

\multicolumn{1}{|c|}{\textbf{RX Peak SNR (dB)}} 
& \multicolumn{1}{c|}{\textbf{57.51}} 
& \multicolumn{1}{c|}{58.9} 
& \multicolumn{1}{c|}{N/A}
& \multicolumn{1}{c|}{49.8}
& \multicolumn{1}{c|}{52.3}
& \multicolumn{1}{c|}{54}
& \multicolumn{1}{c|}{59.4}
\\\hline

\multicolumn{1}{|c|}{\textbf{RX Area/CH $\mathbf{(mm^2)}$}} 
& \multicolumn{1}{c|}{\textbf{0.118$^{*}$}} 
& \multicolumn{1}{c|}{0.065} 
& \multicolumn{1}{c|}{N/A}
& \multicolumn{1}{c|}{0.023}
& \multicolumn{1}{c|}{0.0265}
& \multicolumn{1}{c|}{0.048}
& \multicolumn{1}{c|}{0.0625$^\$$}
\\\hline

\multicolumn{1}{|c|}{\textbf{RX Power/CH (mW)}} 
& \multicolumn{1}{c|}{\textbf{5.83$^{*}$}} 
& \multicolumn{1}{c|}{22.7} 
& \multicolumn{1}{c|}{0.4}
& \multicolumn{1}{c|}{1.54}
& \multicolumn{1}{c|}{0.98}
& \multicolumn{1}{c|}{1.83}
& \multicolumn{1}{c|}{0.95}
\\\hline


\parbox[c][0.8cm]{100pt}{\centering
{\textbf{Output Data Reduction Technique}}} &
\parbox[c][0.8cm]{52pt}{\centering
{\textbf{Compressive Sensing}}}&
\parbox[c][0.8cm]{50pt} {\centering{Digital Beamforming}}&
\parbox[c][0.8cm]{50pt} {\centering{No}} &
\parbox[c][0.8cm]{50pt} {\centering{No}}&
\parbox[c][0.8cm]{50pt} {\centering{Analog Beamforming}}  &
\parbox[c][0.8cm]{50pt} {\centering{Analog Beamforming}} &
\parbox[c][0.8cm]{50pt} {\centering{No}}
\\\hline

\parbox[c][0.8cm]{100pt}{\centering
{\textbf{Data type}}} &
\parbox[c][0.8cm]{52pt}{\centering
{\textbf{Reconstructed full array data}}}&
\parbox[c][0.8cm]{50pt} {\centering{Beamformed-only data}}&
\parbox[c][0.8cm]{50pt} {\centering{Raw full array data}} &
\parbox[c][0.8cm]{50pt} {\centering{Raw full array data}}&
\parbox[c][0.8cm]{50pt} {\centering{Beamformed-only data}}  &
\parbox[c][0.8cm]{50pt} {\centering{Beamformed-only data}} &
\parbox[c][0.8cm]{50pt} {\centering{Raw full array data}}
\\\hline

\end{tabular}

\begin{tablenotes}
\footnotesize
\item[]
$^{*}$ Includes LDO; $^\$$ Includes ultrasound TX.
\end{tablenotes}
\label{tab1}
\end{table*}

\section{Conclusion}
In conclusion, this paper presents a compressive sensing photoacoustic imaging RX architecture that tackles the critical challenge of output data bandwidth and power consumption. By integrating analog-domain spatial compression with an MVM SAR ADC, the RX achieves up to 8x output data rate reduction and 4x ADC count reduction. The MVM SAR ADC performs accurate and fully passive matrix multiplication using ternary-weighted inputs, enabling compact and energy-efficient implementation. The receiver integrates 16 AFEs, achieving an input-referred noise floor of 3.5 nV/$\sqrt{Hz}$ and an ADC's SNDR of 57.5 dB at 20.41 MS/s. Two signal reconstruction methods — FISTA and INR — are implemented to recover from compressed measurements. Phantom imaging experiments with human hairs and a 3D-printed object validate the system’s ability to perform image reconstruction under compression ratios of up to 8 times. While our optimization and demonstration are focused on PA imaging with lower input SNR, the same concepts can be applied to ultrasound imaging systems.  Our prototype with a 65nm RX chip demonstrates the feasibility of embedding compressive sensing into RX hardware for next-generation wearable acoustic imaging systems. 




\ifCLASSOPTIONcaptionsoff
  \newpage
\fi

\bibliography{bib/JSSC_PA_2025.bib,bib/bstcontrol}

@inproceedings{sawaby2018wireless,
  title={A wireless implantable ultrasound array receiver for thermoacoustic imaging},
  author={Sawaby, Ahmed and Wang, Max L and So, Ernest and Chien, Jun-Chau and Nan, Hao and Khuri-Yakub, Butrus T and Arbabian, Amin},
  booktitle={2018 IEEE Symposium on VLSI Circuits},
  pages={189--190},
  year={2018},
  organization={IEEE}
}

@InProceedings{Reiser_2021_ICCV,
    author    = {Reiser, Christian and Peng, Songyou and Liao, Yiyi and Geiger, Andreas},
    title     = {KiloNeRF: Speeding Up Neural Radiance Fields With Thousands of Tiny MLPs},
    booktitle = {Proceedings of the IEEE/CVF International Conference on Computer Vision (ICCV)},
    month     = {October},
    year      = {2021},
    pages     = {14335-14345}
}

@article{miller2017deep,
  title={Deep tissue imaging with multiphoton fluorescence microscopy},
  author={Miller, David R and Jarrett, Jeremy W and Hassan, Ahmed M and Dunn, Andrew K},
  journal={Current opinion in biomedical engineering},
  volume={4},
  pages={32--39},
  year={2017},
  publisher={Elsevier}
}

@inproceedings{saragadam2022miner,
              title={MINER: Multiscale Implicit Neural Representations},
              author={Vishwanath Saragadam and Jasper Tan and Guha Balakrishnan and Richard Baraniuk and Ashok Veeraraghavan},
              booktitle={European Conf. Computer Vision},
              year={2022},
              url_Paper={https://arxiv.org/pdf/2202.03532.pdf}
}

@article{beard2011biomedical,
  title={Biomedical photoacoustic imaging},
  author={Beard, Paul C.},
  journal={Interface Focus},
  volume={1},
  number={4},
  pages={602--631},
  year={2011},
  publisher={The Royal Society},
  doi={10.1098/rsfs.2011.0028},
  url={https://doi.org/10.1098/rsfs.2011.0028}
}

@article{guo_fully_2017,
	title = {A {Fully} {Passive} {Compressive} {Sensing} {SAR} {ADC} for {Low}-{Power} {Wireless} {Sensors}},
	volume = {52},
	issn = {1558-173X},
	doi = {10.1109/JSSC.2017.2695573},
	number = {8},
	journal = {IEEE Journal of Solid-State Circuits},
	author = {Guo, Wenjuan and Kim, Youngchun and Tewfik, Ahmed H. and Sun, Nan},
	month = aug,
	year = {2017},
	pages = {2154--2167},
}

@inproceedings{bindra_174vrms_2022,
	title = {A {174~$\mu$VRMS} {Input} {Noise}, 1 {G8}/s {Comparator} in 22nm {FDSOI} with a {Dynamic}-{Bias} {Preamplifier} {Using} {Tail} {Charge} {Pump} and {Capacitive} {Neutralization} {Across} the {Latch}},
	volume = {65},
	doi = {10.1109/ISSCC42614.2022.9731728},
	booktitle = {2022 {IEEE} {International} {Solid}- {State} {Circuits} {Conference} ({ISSCC})},
	author = {Bindra, Harijot Singh and Ponte, Jeroen and Nauta, Bram},
	month = feb,
	year = {2022},
	pages = {1--3},
}

@article{chen_column-row-parallel_2016,
	title = {A {Column}-{Row}-{Parallel} {ASIC} {Architecture} for 3-{D} {Portable} {Medical} {Ultrasonic} {Imaging}},
	volume = {51},
	issn = {1558-173X},
	doi = {10.1109/JSSC.2015.2505714},
	number = {3},
	journal = {IEEE Journal of Solid-State Circuits},
	author = {Chen, Kailiang and Lee, Hae-Seung and Sodini, Charles G.},
	month = mar,
	year = {2016},	pages = {738--751},
}

@article{chen_pixel_2017,
	title = {A {Pixel} {Pitch}-{Matched} {Ultrasound} {Receiver} for 3-{D} {Photoacoustic} {Imaging} {With} {Integrated} {Delta}-{Sigma} {Beamformer} in 28-nm {UTBB} {FD}-{SOI}},
	volume = {52},
	issn = {1558-173X},
	doi = {10.1109/JSSC.2017.2749425},
	number = {11},
	journal = {IEEE Journal of Solid-State Circuits},
	author = {Chen, Man-Chia and Peña Perez, Aldo and Kothapalli, Sri-Rajasekhar and Cathelin, Philippe and Cathelin, Andreia and Gambhir, Sanjiv Sam and Murmann, Boris},
	month = nov,
	year = {2017},	pages = {2843--2856},
}

@article{lee_36-channel_2021,
	title = {A 36-{Channel} {Auto}-{Calibrated} {Front}-{End} {ASIC} for a {pMUT}-{Based} {Miniaturized} 3-{D} {Ultrasound} {System}},
	volume = {56},
	issn = {1558-173X},
	doi = {10.1109/JSSC.2021.3049560},
	number = {6},
	journal = {IEEE Journal of Solid-State Circuits},
	author = {Lee, Jihee and Lee, Kyoung-Rog and Eovino, Benjamin E. and Park, Jeong Hoan and Liang, Luna Yue and Lin, Liwei and Yoo, Hoi-Jun and Yoo, Jerald},
	month = jun,
	year = {2021},
	pages = {1910--1923},
}

@article{hopf_pitch-matched_2023,
	title = {A {Pitch}-{Matched} {High}-{Frame}-{Rate} {Ultrasound} {Imaging} {ASIC} for {Catheter}-{Based} 3-{D} {Probes}},
	issn = {1558-173X},
	doi = {10.1109/JSSC.2023.3299749},
	journal = {IEEE Journal of Solid-State Circuits},
	author = {Hopf, Yannick M. and dos Santos, Djalma Simoes and Ossenkoppele, Boudewine W. and Soozande, Mehdi and Noothout, Emile and Chang, Zu-Yao and Chen, Chao and Vos, Hendrik J. and Bosch, Johan G. and Verweij, Martin D. and de Jong, Nico and Pertijs, Michiel A. P.},
	year = {2023},	pages = {1--16},
}

@book{hoskins2019diagnostic,
  title={Diagnostic ultrasound: physics and equipment},
  author={Hoskins, Peter and Martin, Kevin and Thrush, Abigail},
  year={2019},
  publisher={CRC Press}
}

@article{ermilov2009laser,
  title={Laser optoacoustic imaging system for detection of breast cancer},
  author={Ermilov, Sergey A and Khamapirad, Tuenchit and Conjusteau, Andre and Leonard, Morton H and Lacewell, Ron and Mehta, Ketan and Miller, Tom and Oraevsky, Alexander A},
  journal={Journal of biomedical optics},
  volume={14},
  number={2},
  pages={024007--024007},
  year={2009},
  publisher={Society of Photo-Optical Instrumentation Engineers}
}

@article{zhu_towards_2020,
	title = {Towards {Clinical} {Translation} of {LED}-{Based} {Photoacoustic} {Imaging}: {A} {Review}},
	volume = {20},
	copyright = {http://creativecommons.org/licenses/by/3.0/},
	issn = {1424-8220},
	shorttitle = {Towards {Clinical} {Translation} of {LED}-{Based} {Photoacoustic} {Imaging}},
	url = {https://www.mdpi.com/1424-8220/20/9/2484},
	doi = {10.3390/s20092484},
	language = {en},
	number = {9},
	urldate = {2024-10-28},
	journal = {Sensors},
	author = {Zhu, Yunhao and Feng, Ting and Cheng, Qian and Wang, Xueding and Du, Sidan and Sato, Naoto and Yuan, Jie and Kuniyil Ajith Singh, Mithun},
	month = jan,
	year = {2020},
	note = {Number: 9
Publisher: Multidisciplinary Digital Publishing Institute},
	pages = {2484},
}

@article{zhu_light_2018,
	title = {Light {Emitting} {Diodes} based {Photoacoustic} {Imaging} and {Potential} {Clinical} {Applications}},
	volume = {8},
	copyright = {2018 The Author(s)},
	issn = {2045-2322},
	url = {https://www.nature.com/articles/s41598-018-28131-4},
	doi = {10.1038/s41598-018-28131-4},
	language = {en},
	number = {1},
	urldate = {2024-10-28},
	journal = {Scientific Reports},
	author = {Zhu, Yunhao and Xu, Guan and Yuan, Jie and Jo, Janggun and Gandikota, Girish and Demirci, Hakan and Agano, Toshitaka and Sato, Naoto and Shigeta, Yusuke and Wang, Xueding},
	month = jun,
	year = {2018},
	pages = {9885},
}

@inproceedings{li_154mwelement_2019,
	title = {A 1.{54mW}/{Element} 150~$\mu$m-{Pitch}-{Matched} {Receiver} {ASIC} with {Element}-{Level} {SAR}/{Shared}-{Single}-{Slope} {Hybrid} {ADCs} for {Miniature} {3D} {Ultrasound} {Probes}},
	url = {https://ieeexplore.ieee.org/abstract/document/8778200},
	doi = {10.23919/VLSIC.2019.8778200},
	urldate = {2023-12-19},
	booktitle = {2019 {Symposium} on {VLSI} {Circuits}},
	author = {Li, Jing and Chen, Zhao and Tan, Mingliang and van Willigen, Douwe and Chen, Chao and Chang, Zu-yao and Noothout, Emile and de Jong, Nico and Verweij, Martin and Pertijs, Michiel},
	month = jun,
	year = {2019},
	pages = {C220--C221},
}

@inproceedings{hopf_pitch-matched_2022,
	address = {San Francisco, CA, USA},
	title = {A {Pitch}-{Matched} {ASIC} with {Integrated} {65V} {TX} and {Shared} {Hybrid} {Beamforming} {ADC} for {Catheter}-{Based} {High}-{Frame}-{Rate} {3D} {Ultrasound} {Probes}},
	isbn = {978-1-6654-2800-2},
	url = {https://ieeexplore.ieee.org/document/9731597/},
	doi = {10.1109/ISSCC42614.2022.9731597},
	language = {en},
	urldate = {2023-12-19},
	booktitle = {2022 {IEEE} {International} {Solid}- {State} {Circuits} {Conference} ({ISSCC})},
	publisher = {IEEE},
	author = {Hopf, Yannick and Ossenkoppele, Boudewine and Soozande, Mehdi and Noothout, Emile and Chang, Zu-Yao and Chen, Chao and Vos, Hendrik and Bosch, Hans and Verweij, Martin and De Jong, Nico and Pertijs, Michiel},
	month = feb,
	year = {2022},
	pages = {494--496},
}

@article{guo_125_2024,
	title = {A 125~$\mu$m-{Pitch}-{Matched} {Transceiver} {ASIC} {With} {Micro}-{Beamforming} {ADC} and {Multi}-{Level} {Signaling} for 3-{D} {Transfontanelle} {Ultrasonography}},
	issn = {1558-173X},
	url = {https://ieeexplore.ieee.org/abstract/document/10418877},
	doi = {10.1109/JSSC.2024.3355854},
	urldate = {2024-02-08},
	journal = {IEEE Journal of Solid-State Circuits},
	author = {Guo, Peng and Fool, Fabian and Chang, Zu-Yao and Noothout, Emile and Vos, Hendrik J. and Bosch, Johan G. and de Jong, Nico and Verweij, Martin D. and Pertijs, Michiel A. P.},
	year = {2024},
	pages = {1--14},
}

@inproceedings{yoo_100mhz2ghz_2012,
	address = {Montreal, QC, Canada},
	title = {A {100MHz}–{2GHz} 12.5x sub-{Nyquist} rate receiver in 90nm {CMOS}},
	isbn = {978-1-4673-0413-9 978-1-4673-0416-0},
	url = {https://ieeexplore.ieee.org/document/6242225/},
	doi = {10.1109/RFIC.2012.6242225},
	language = {en},
	urldate = {2024-08-20},
	booktitle = {2012 {IEEE} {Radio} {Frequency} {Integrated} {Circuits} {Symposium}},
	publisher = {IEEE},
	author = {Yoo, Juhwan and Becker, Stephen and Loh, Matthew and Monge, Manuel and Candes, Emmanuel and Emami-Neyestanak, Azita},
	month = jun,
	year = {2012},
	pages = {31--34},
}

@article{soozande_imaging_2022,
	title = {Imaging {Scheme} for 3-{D} {High}-{Frame}-{Rate} {Intracardiac} {Echography}: {A} {Simulation} {Study}},
	volume = {69},
	issn = {1525-8955},
	shorttitle = {Imaging {Scheme} for 3-{D} {High}-{Frame}-{Rate} {Intracardiac} {Echography}},
	url = {https://ieeexplore.ieee.org/document/9807327/},
	doi = {10.1109/TUFFC.2022.3186487},
	number = {10},
	urldate = {2024-08-22},
	journal = {IEEE Transactions on Ultrasonics, Ferroelectrics, and Frequency Control},
	author = {Soozande, Mehdi and Ossenkoppele, Boudewine W. and Hopf, Yannick and Pertijs, Michiel A. P. and Verweij, Martin D. and de Jong, Nico and Vos, Hendrik J. and Bosch, Johan G.},
	month = oct,
	year = {2022},	pages = {2862--2874},
}

@article{hu_wearable_2023,
	title = {A wearable cardiac ultrasound imager},
	volume = {613},
	issn = {0028-0836, 1476-4687},
	url = {https://www.nature.com/articles/s41586-022-05498-z},
	doi = {10.1038/s41586-022-05498-z},
	language = {en},
	number = {7945},
	urldate = {2024-08-30},
	journal = {Nature},
	author = {Hu, Hongjie and Huang, Hao and Li, Mohan and Gao, Xiaoxiang and Yin, Lu and Qi, Ruixiang and Wu, Ray S. and Chen, Xiangjun and Ma, Yuxiang and Shi, Keren and Li, Chenghai and Maus, Timothy M. and Huang, Brady and Lu, Chengchangfeng and Lin, Muyang and Zhou, Sai and Lou, Zhiyuan and Gu, Yue and Chen, Yimu and Lei, Yusheng and Wang, Xinyu and Wang, Ruotao and Yue, Wentong and Yang, Xinyi and Bian, Yizhou and Mu, Jing and Park, Geonho and Xiang, Shu and Cai, Shengqiang and Corey, Paul W. and Wang, Joseph and Xu, Sheng},
	month = jan,
	year = {2023},
	pages = {667--675},
}

@article{ryu_comprehensive_2021,
	title = {Comprehensive pregnancy monitoring with a network of wireless, soft, and flexible sensors in high- and low-resource health settings},
	volume = {118},
	issn = {0027-8424, 1091-6490},
	url = {https://pnas.org/doi/full/10.1073/pnas.2100466118},
	doi = {10.1073/pnas.2100466118},
	language = {en},
	number = {20},
	urldate = {2024-08-30},
	journal = {Proceedings of the National Academy of Sciences},
	author = {Ryu, Dennis and Kim, Dong Hyun and Price, Joan T. and Lee, Jong Yoon and Chung, Ha Uk and Allen, Emily and Walter, Jessica R. and Jeong, Hyoyoung and Cao, Jingyue and Kulikova, Elena and Abu-Zayed, Hajar and Lee, Rachel and Martell, Knute L. and Zhang, Michael and Kampmeier, Brianna R. and Hill, Marc and Lee, JooHee and Kim, Edward and Park, Yerim and Jang, Hokyung and Arafa, Hany and Liu, Claire and Chisembele, Maureen and Vwalika, Bellington and Sindano, Ntazana and Spelke, M. Bridget and Paller, Amy S. and Premkumar, Ashish and Grobman, William A. and Stringer, Jeffrey S. A. and Rogers, John A. and Xu, Shuai},
	month = may,
	year = {2021},
	pages = {e2100466118},
}

@article{protopappas_ultrasound_2005,
	title = {An ultrasound wearable system for the monitoring and acceleration of fracture healing in long bones},
	volume = {52},
	issn = {1558-2531},
	url = {https://ieeexplore.ieee.org/document/1495704/?arnumber=1495704},
	doi = {10.1109/TBME.2005.851507},
	number = {9},
	urldate = {2024-08-30},
	journal = {IEEE Transactions on Biomedical Engineering},
	author = {Protopappas, V.C. and Baga, D.A. and Fotiadis, D.I. and Likas, A.C. and Papachristos, A.A. and Malizos, K.N.},
	month = sep,
	year = {2005},	pages = {1597--1608},
}

@article{wang_photoacoustic_2012,
	title = {Photoacoustic {Tomography}: {In} {Vivo} {Imaging} from {Organelles} to {Organs}},
	volume = {335},
	shorttitle = {Photoacoustic {Tomography}},
	url = {https://www.science.org/doi/full/10.1126/science.1216210},
	doi = {10.1126/science.1216210},
	number = {6075},
	urldate = {2024-08-30},
	journal = {Science},
	author = {Wang, Lihong V. and Hu, Song},
	month = mar,
	year = {2012},	pages = {1458--1462},
}

@article{xu_photoacoustic_2006,
	title = {Photoacoustic imaging in biomedicine},
	volume = {77},
	issn = {0034-6748, 1089-7623},
	url = {https://pubs.aip.org/rsi/article/77/4/041101/913562/Photoacoustic-imaging-in-biomedicine},
	doi = {10.1063/1.2195024},
	language = {en},
	number = {4},
	urldate = {2024-09-02},
	journal = {Review of Scientific Instruments},
	author = {Xu, Minghua and Wang, Lihong V.},
	month = apr,
	year = {2006},
	pages = {041101},
}

@article{farrell_coir_2023,
	title = {{CoIR}: {Compressive} {Implicit} {Radar}},
	issn = {1939-3539},
	shorttitle = {{CoIR}},
	url = {https://ieeexplore.ieee.org/abstract/document/10214469?casa_token=wPNe-sxrTPYAAAAA:FdHh7bDcOUyKUqoFu-g72MF-XqwgXykUnNVrkaS7_6c5Ydhap06SjgMkoea5gwGRZ8SsZ5uJ},
	doi = {10.1109/TPAMI.2023.3301553},
	urldate = {2024-09-02},
	journal = {IEEE Transactions on Pattern Analysis and Machine Intelligence},
	author = {Farrell, Sean M. and Boominathan, Vivek and Raymondi, Nathaniel and Sabharwal, Ashutosh and Veeraraghavan, Ashok},
	year = {2023},	pages = {1--12},
}

@inproceedings{sitzmann_implicit_2020,
	title = {Implicit {Neural} {Representations} with {Periodic} {Activation} {Functions}},
	volume = {33},
	url = {https://proceedings.neurips.cc/paper/2020/hash/53c04118df112c13a8c34b38343b9c10-Abstract.html},
	urldate = {2024-09-02},
	booktitle = {Advances in {Neural} {Information} {Processing} {Systems}},
	publisher = {Curran Associates, Inc.},
	author = {Sitzmann, Vincent and Martel, Julien and Bergman, Alexander and Lindell, David and Wetzstein, Gordon},
	year = {2020},
	pages = {7462--7473},
}

@article{beck_fast_2009,
	title = {A {Fast} {Iterative} {Shrinkage}-{Thresholding} {Algorithm} for {Linear} {Inverse} {Problems}},
	volume = {2},
	issn = {1936-4954},
	url = {http://epubs.siam.org/doi/10.1137/080716542},
	doi = {10.1137/080716542},
	language = {en},
	number = {1},
	urldate = {2024-09-03},
	journal = {SIAM Journal on Imaging Sciences},
	author = {Beck, Amir and Teboulle, Marc},
	month = jan,
	year = {2009},
	pages = {183--202},
}

@article{kruizinga_compressive_2017,
	title = {Compressive {3D} ultrasound imaging using a single sensor},
	volume = {3},
	issn = {2375-2548},
	url = {https://www.science.org/doi/10.1126/sciadv.1701423},
	doi = {10.1126/sciadv.1701423},
	language = {en},
	number = {12},
	urldate = {2024-09-16},
	journal = {Science Advances},
	author = {Kruizinga, Pieter and Van Der Meulen, Pim and Fedjajevs, Andrejs and Mastik, Frits and Springeling, Geert and De Jong, Nico and Bosch, Johannes G. and Leus, Geert},
	month = dec,
	year = {2017},
	pages = {e1701423},
}

@article{shoaran_compact_2014,
	title = {Compact {Low}-{Power} {Cortical} {Recording} {Architecture} for {Compressive} {Multichannel} {Data} {Acquisition}},
	volume = {8},
	issn = {1940-9990},
	url = {https://ieeexplore.ieee.org/document/6786004/?arnumber=6786004},
	doi = {10.1109/TBCAS.2014.2304582},
	number = {6},
	urldate = {2024-10-14},
	journal = {IEEE Transactions on Biomedical Circuits and Systems},
	author = {Shoaran, Mahsa and Kamal, Mahdad Hosseini and Pollo, Claudio and Vandergheynst, Pierre and Schmid, Alexandre},
	month = dec,
	year = {2014},	pages = {857--870},
}

@article{park_clinical_2024,
	title = {Clinical translation of photoacoustic imaging},
	issn = {2731-6092},
	url = {https://www.nature.com/articles/s44222-024-00240-y},
	doi = {10.1038/s44222-024-00240-y},
	language = {en},
	urldate = {2025-01-06},
	journal = {Nature Reviews Bioengineering},
	author = {Park, Jeongwoo and Choi, Seongwook and Knieling, Ferdinand and Clingman, Bryan and Bohndiek, Sarah and Wang, Lihong V. and Kim, Chulhong},
	month = sep,
	year = {2024},
}

@inproceedings{liao_352_2025,
	title = {35.2 {A} {Spatial}-{Domain} {Compressive}-{Sensing} {Photoacoustic} {Imager} with {Matrix}-{Multiplying} {SAR} {ADC}},
	volume = {68},
	url = {https://ieeexplore.ieee.org/abstract/document/10904572},
	doi = {10.1109/ISSCC49661.2025.10904572},
	urldate = {2025-05-01},
	booktitle = {2025 {IEEE} {International} {Solid}-{State} {Circuits} {Conference} ({ISSCC})},
	author = {Liao, Huan-Cheng and Zhang, Shunyao and Su, Yumin and Govinday, Arvind and Zou, Yiwei and Wang, Wei and Boominathan, Vivek and Veeraraghavan, Ashok and Li, Lei and Yang, Kaiyuan},
	month = feb,
	year = {2025},	pages = {1--3},
}

@article{harpe_26_2011,
	title = {A 26 ~$\mu$ {W} 8 bit 10 {MS}/s {Asynchronous} {SAR} {ADC} for {Low} {Energy} {Radios}},
	volume = {46},
	issn = {1558-173X},
	url = {https://ieeexplore.ieee.org/document/5771068/?arnumber=5771068},
	doi = {10.1109/JSSC.2011.2143870},
	number = {7},
	urldate = {2025-04-05},
	journal = {IEEE Journal of Solid-State Circuits},
	author = {Harpe, Pieter J. A. and Zhou, Cui and Bi, Yu and van der Meijs, Nick P. and Wang, Xiaoyan and Philips, Kathleen and Dolmans, Guido and de Groot, Harmke},
	month = jul,
	year = {2011},	pages = {1585--1595},
}

@article{donoho_compressed_2006,
	title = {Compressed sensing},
	volume = {52},
	copyright = {https://ieeexplore.ieee.org/Xplorehelp/downloads/license-information/IEEE.html},
	issn = {0018-9448},
	url = {http://ieeexplore.ieee.org/document/1614066/},
	doi = {10.1109/TIT.2006.871582},
	number = {4},
	urldate = {2025-05-06},
	journal = {IEEE Transactions on Information Theory},
	author = {Donoho, D.L.},
	month = apr,
	year = {2006},
	pages = {1289--1306},
}

@article{baraniuk_compressive_2007,
	title = {Compressive {Sensing} [{Lecture} {Notes}]},
	volume = {24},
	copyright = {https://ieeexplore.ieee.org/Xplorehelp/downloads/license-information/IEEE.html},
	issn = {1053-5888},
	url = {http://ieeexplore.ieee.org/document/4286571/},
	doi = {10.1109/MSP.2007.4286571},
	number = {4},
	urldate = {2025-05-06},
	journal = {IEEE Signal Processing Magazine},
	author = {Baraniuk, Richard},
	month = jul,
	year = {2007},
	pages = {118--121},
}

@article{candes_sparsity_2007,
	title = {Sparsity and incoherence in compressive sampling},
	volume = {23},
	issn = {0266-5611, 1361-6420},
	url = {https://iopscience.iop.org/article/10.1088/0266-5611/23/3/008},
	doi = {10.1088/0266-5611/23/3/008},
	number = {3},
	urldate = {2025-05-06},
	journal = {Inverse Problems},
	author = {Candès, Emmanuel and Romberg, Justin},
	month = jun,
	year = {2007},
	pages = {969--985},
}

@article{candes_decoding_2005,
	title = {Decoding by linear programming},
	volume = {51},
	issn = {1557-9654},
	url = {https://ieeexplore.ieee.org/document/1542412/},
	doi = {10.1109/TIT.2005.858979},
	number = {12},
	urldate = {2025-05-07},
	journal = {IEEE Transactions on Information Theory},
	author = {Candes, E.J. and Tao, T.},
	month = dec,
	year = {2005},
	pages = {4203--4215},
}

@article{stuker_fluorescence_2011,
	title = {Fluorescence {Molecular} {Tomography}: {Principles} and {Potential} for {Pharmaceutical} {Research}},
	volume = {3},
	copyright = {http://creativecommons.org/licenses/by/3.0/},
	issn = {1999-4923},
	shorttitle = {Fluorescence {Molecular} {Tomography}},
	url = {https://www.mdpi.com/1999-4923/3/2/229},
	doi = {10.3390/pharmaceutics3020229},
	language = {en},
	number = {2},
	urldate = {2025-05-08},
	journal = {Pharmaceutics},
	author = {Stuker, Florian and Ripoll, Jorge and Rudin, Markus},
	month = jun,
	year = {2011},	pages = {229--274},
}

@article{kodach_quantitative_2010,
	title = {Quantitative comparison of the {OCT} imaging depth at 1300 nm and 1600 nm},
	volume = {1},
	copyright = {© 2010 OSA},
	issn = {2156-7085},
	url = {https://opg.optica.org/boe/abstract.cfm?uri=boe-1-1-176},
	doi = {10.1364/BOE.1.000176},
	language = {EN},
	number = {1},
	urldate = {2025-05-08},
	journal = {Biomedical Optics Express},
	author = {Kodach, V. M. and Kalkman, J. and Faber, D. J. and Leeuwen, T. G. van},
	month = aug,
	year = {2010},	pages = {176--185},
}

@article{kim_deeply_2010,
	title = {Deeply penetrating in vivo photoacoustic imaging using a clinical ultrasound array system},
	volume = {1},
	copyright = {© 2010 OSA},
	issn = {2156-7085},
	url = {https://opg.optica.org/boe/abstract.cfm?uri=boe-1-1-278},
	doi = {10.1364/BOE.1.000278},
	language = {EN},
	number = {1},
	urldate = {2025-05-09},
	journal = {Biomedical Optics Express},
	author = {Kim, Chulhong and Erpelding, Todd N. and Jankovic, Ladislav and Pashley, Michael D. and Wang, Lihong V.},
	month = aug,
	year = {2010},	pages = {278--284},
}

@article{siphanto_serial_2005,
	title = {Serial noninvasive photoacoustic imaging of neovascularization in tumor angiogenesis},
	volume = {13},
	copyright = {© 2005 Optical Society of America},
	issn = {1094-4087},
	url = {https://opg.optica.org/oe/abstract.cfm?uri=oe-13-1-89},
	doi = {10.1364/OPEX.13.000089},
	language = {EN},
	number = {1},
	urldate = {2025-05-09},
	journal = {Optics Express},
	author = {Siphanto, R. I. and Thumma, K. K. and Kolkman, R. G. M. and Leeuwen, T. G. van and Mul, F. F. M. de and Neck, J. W. van and Adrichem, L. N. A. van and Steenbergen, W.},
	month = jan,
	year = {2005},	pages = {89--95},
}

@article{gurun_analog_2012,
	title = {An {Analog} {Integrated} {Circuit} {Beamformer} for {High}-{Frequency} {Medical} {Ultrasound} {Imaging}},
	volume = {6},
	issn = {1940-9990},
	url = {https://ieeexplore.ieee.org/abstract/document/6339020},
	doi = {10.1109/TBCAS.2012.2219532},
	number = {5},
	urldate = {2025-05-09},
	journal = {IEEE Transactions on Biomedical Circuits and Systems},
	author = {Gurun, Gokce and Zahorian, Jaime S. and Sisman, Alper and Karaman, Mustafa and Hasler, Paul E. and Degertekin, F. Levent},
	month = oct,
	year = {2012},
	pages = {454--467},
}

@inproceedings{cao_103_2023,
	title = {10.3 {A} {Single}-{Channel} 12b {2GS}/s {PVT}-{Robust} {Pipelined} {ADC} with {Critically} {Damped} {Ring} {Amplifier} and {Time}-{Domain} {Quantizer}},
	url = {https://ieeexplore.ieee.org/document/10067687/},
	doi = {10.1109/ISSCC42615.2023.10067687},
	urldate = {2025-05-09},
	booktitle = {2023 {IEEE} {International} {Solid}-{State} {Circuits} {Conference} ({ISSCC})},
	author = {Cao, Yuefeng and Zhang, Minglei and Zhu, Yan and Chan, Chi-Hang and Martins, R. P.},
	month = feb,
	year = {2023},	pages = {9--11},
}

@inproceedings{wang_reconfigurable_2018,
	title = {A {Reconfigurable} {Analog} {Baseband} for {Ka} {Band} {Transmitter}},
	url = {https://ieeexplore.ieee.org/document/8487143},
	doi = {10.1109/EDSSC.2018.8487143},
	urldate = {2025-05-09},
	booktitle = {2018 {IEEE} {International} {Conference} on {Electron} {Devices} and {Solid} {State} {Circuits} ({EDSSC})},
	author = {Wang, Wei and Ye, Jialiang and Zheng, Wankai and Chen, Zhiyuan and Liao, YiXiu and Chi, Baoyong},
	month = jun,
	year = {2018},
	pages = {1--2},
}

@article{kim_single-chip_2017,
	title = {A {Single}-{Chip} 64-{Channel} {Ultrasound} {RX}-{Beamformer} {Including} {Analog} {Front}-{End} and an {LUT} for {Non}-{Uniform} {ADC}-{Sample}-{Clock} {Generation}},
	volume = {11},
	issn = {1940-9990},
	url = {https://ieeexplore.ieee.org/document/7546856/},
	doi = {10.1109/TBCAS.2016.2571739},
	number = {1},
	urldate = {2025-05-09},
	journal = {IEEE Transactions on Biomedical Circuits and Systems},
	author = {Kim, Yoon-Jee and Cho, Sung-Eun and Um, Ji-Yong and Chae, Min-Kyun and Bang, Jihoon and Song, Jongkeun and Jeon, Taeho and Kim, Byungsub and Sim, Jae-Yoon and Park, Hong-June},
	month = feb,
	year = {2017},
	pages = {87--97},
}

@article{liu_finer_nodate,
	title = {{FINER}: {Flexible} spectral-bias tuning in {Implicit} {NEural} {Representation} by {Variable}-periodic {Activation} {Functions}: {Supplemental} {Material}},
	language = {en},
	author = {Liu, Zhen and Zhu, Hao and Zhang, Qi and Fu, Jingde and Deng, Weibing and Ma, Zhan and Guo, Yanwen and Cao, Xun},
}

@IEEEtranBSTCTL{IEEEexample:BSTcontrol,
  CTLuse_forced_etal       = "yes",
  CTLmax_names_forced_etal = "6",
  CTLnames_show_etal       = "1",
  CTLuse_url               = "no",
  CTLdash_repeated_names   = "no",
CTLname_url_prefix = "[Online]. Available:"
}
\bibliographystyle{IEEEtran}

\begin{IEEEbiography}[{\includegraphics[width=1in,height=1.25in,clip,keepaspectratio]{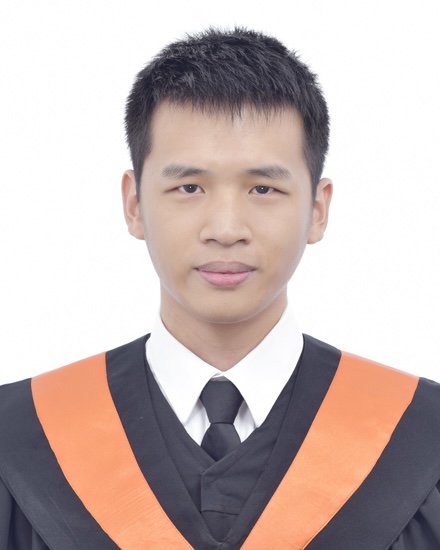}}]{Huan-Cheng Liao}
(Graduate Student Member, IEEE) received the B.S. degree in Engineering Science from National Taiwan University, Taiwan, in 2020. He is currently working toward his Ph.D. degree in Electrical and Computer Engineering at Rice University, Houston, TX, USA. His research interests include analog and mixed-signal integrated circuits design.
\end{IEEEbiography}

\vskip -1\baselineskip
\begin{IEEEbiography}[{\includegraphics[width=1in,height=1.25in,clip,keepaspectratio]{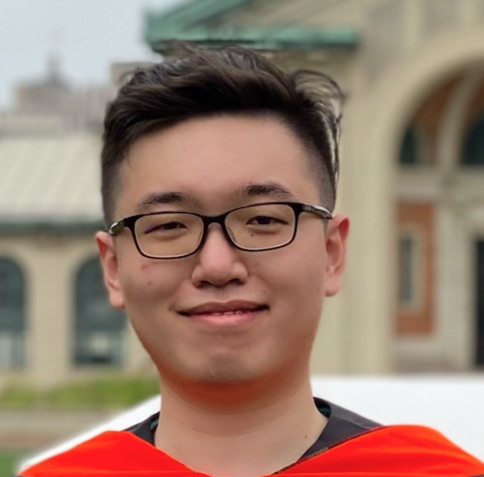}}]{Shunyao Zhang} is a PhD student at Rice University, Houston, Texas, United States, instructed by Prof. Lei S. Li. He received his master’s degree in electrical and computer engineering from Carnegie Mellon University, Pittsburgh, Pennsylvania, United States. His research interests are photoacoustic imaging and deep learning.
\end{IEEEbiography}

\vskip -1\baselineskip
\begin{IEEEbiography}[{\includegraphics[width=1in,height=1.25in,clip,keepaspectratio]{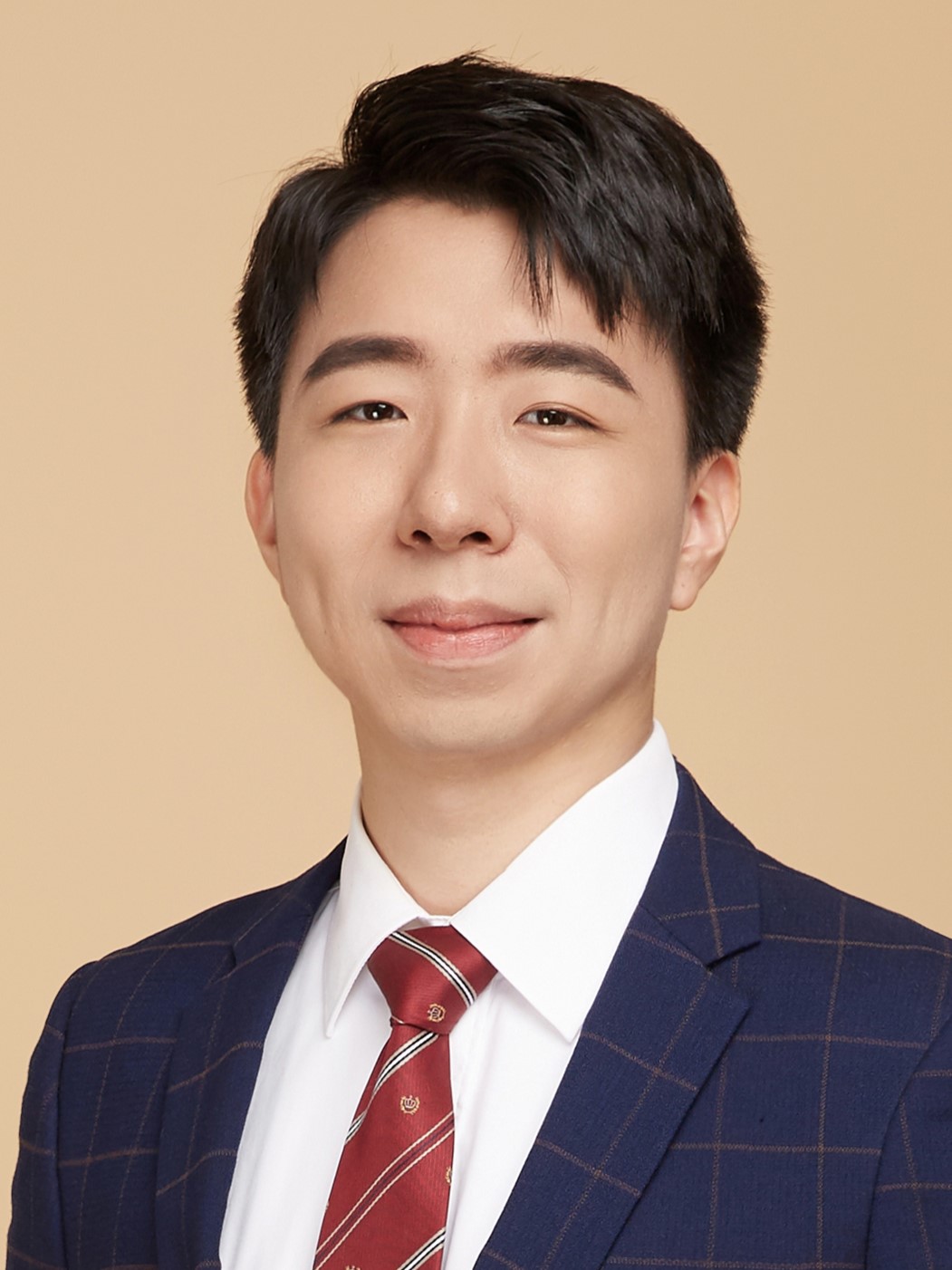}}]{Yumin Su} (Graduate Student Member, IEEE) received his B.S. degree in Electrical and Computer Engineering from Rice University, Houston, TX, USA, in 2023. He graduated summa cum laude with a Distinction in Research and Creative Work. He is currently a 2nd-year Electrical and Computer Engineering Ph.D. student at Rice University, advised by Prof. Kaiyuan Yang. His research interests include low-cost hardware security and design automation. 
\end{IEEEbiography}

\vskip -1\baselineskip
\begin{IEEEbiography}[{\includegraphics[width=1in,height=1.25in,clip,keepaspectratio]{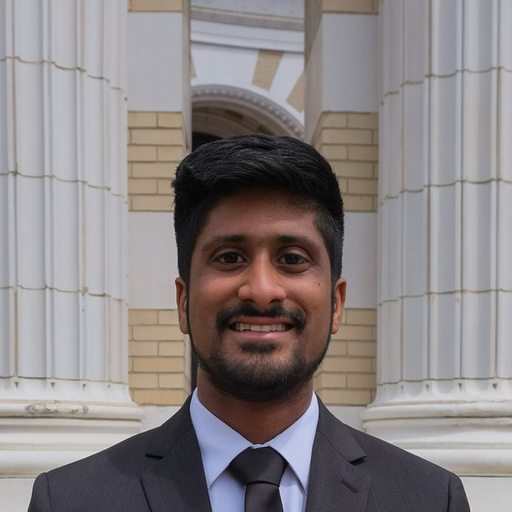}}]{Arvind Govinday} received the B.S. and M.S. degrees in electrical and computer engineering from Carnegie Mellon University, Pittsburgh, PA, USA in 2022. He joined Rice University, Houston, TX, USA as a graduate student researcher in 2023, where he conducted this research as a part of his work in computational imaging. He is currently a Machine Learning Engineer with Hive AI, San Francisco, CA, USA, where he is the primary developer of deep learning models for object detection that serve major media and entertainment clients.
\end{IEEEbiography}

\vskip -1\baselineskip
\begin{IEEEbiography}
[{\includegraphics[width=1in,height=1.25in,clip,keepaspectratio]{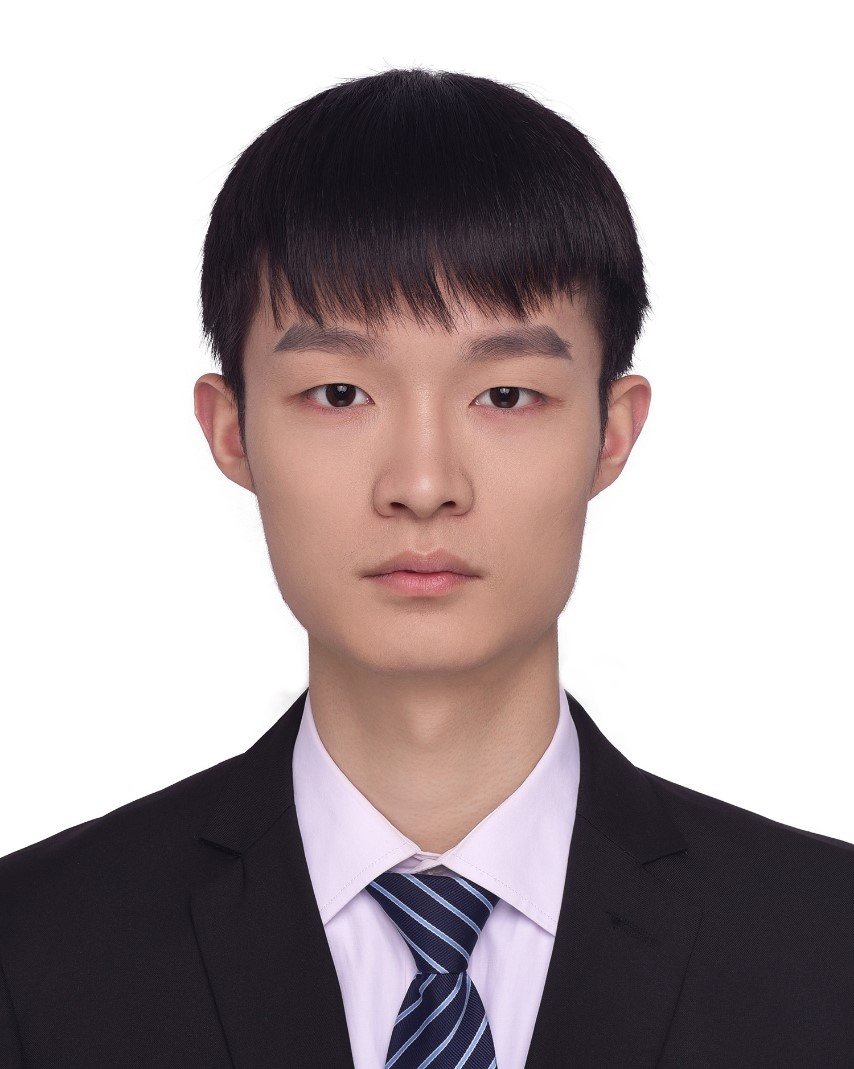}}]{Yiwei Zou} (Graduate Student Member, IEEE) received the B.E. degree in Integrated Circuits and Systems from Huazhong University of Science and Technology, Wuhan, China, in 2022. He is currently working toward his Ph.D. degree in Electrical and Computer Engineering at Rice University, Houston, TX. His research interests include analog and mixed-signal integrated circuits design for bio-electronics.
\end{IEEEbiography}

\vskip -1\baselineskip
\begin{IEEEbiography}
[{\includegraphics[width=1in,height=1.25in,clip,keepaspectratio]{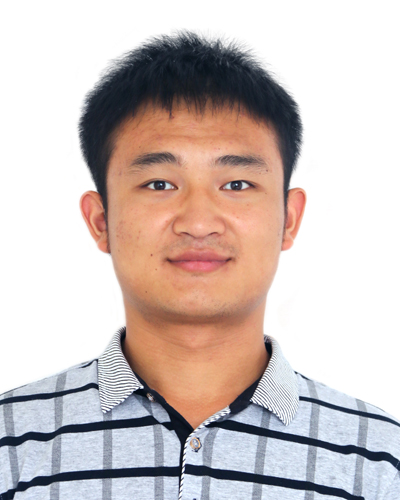}}]{Wei Wang} (Graduate Student Member, IEEE) received a B.S. degree in electronic information science and technology from the Harbin Institute of Technology, Harbin, China, in 2016, and an M.S. degree in integrated circuit engineering from Tsinghua University, Beijing, China, in 2019. He is currently pursuing the Ph.D. degree in electrical and computer engineering with Rice University, Houston, TX, USA. His research interests include mixed-signal circuits and systems design.
\end{IEEEbiography}

\vskip -1\baselineskip
\begin{IEEEbiography}
[{\includegraphics[width=1in,height=1.25in,clip,keepaspectratio]{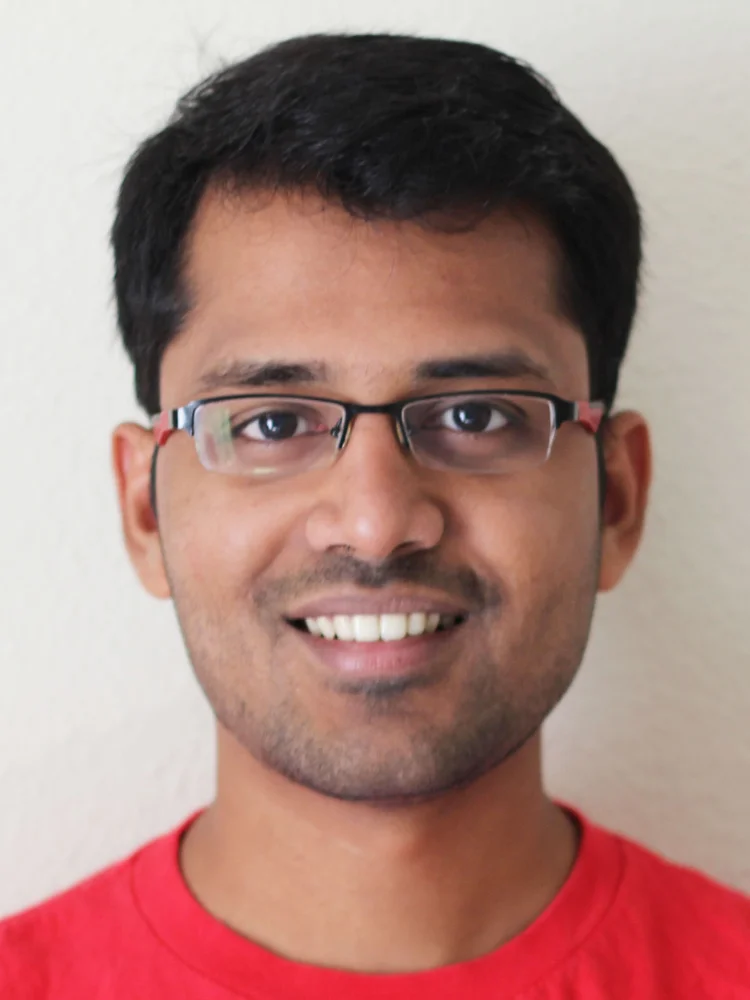}}]{Vivek Boominathan} (Member, IEEE) is an Assistant Research Professor in the Department of Electrical and Computer Engineering at Rice University, Houston, TX. He received his bachelor’s degree in Electrical Engineering from the Indian Institute of Technology, Hyderabad, India, in 2012 and the MS and Ph.D. degrees from the Department of Electrical and Computer Engineering, Rice University, Houston, TX, in 2016 and 2019, respectively. His research interests include computational imaging, computer vision, applied optics, and machine learning.
\end{IEEEbiography}

\vskip -1\baselineskip
\begin{IEEEbiography}
[{\includegraphics[width=1in,height=1.25in,clip,keepaspectratio]{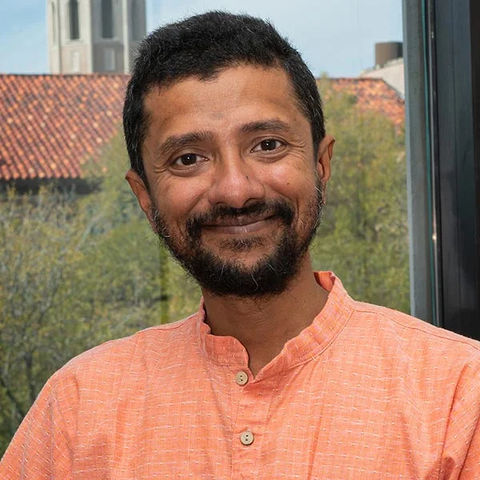}}]{Ashok Veeraraghava} (Fellow, IEEE) received the B.S. degree in electrical engineering from the Indian Institute of Technology, in 2002, and the M.S. and Ph.D. degrees from the Department of Electrical and Computer Engineering, University of Maryland, in 2004 and 2008, respectively. He is currently the Chair and Professor of electrical and computer engineering at Rice University. Before joining Rice University, he spent three years as a Research Scientist with Mitsubishi Electric Research Labs, Cambridge, MA, USA. His research interests include computational imaging, computer vision, machine learning, and robotics. His thesis received the Doctoral Dissertation Award from the Department of Electrical and Computer Engineering at the University of Maryland. He received the NSF CAREER Award in 2017.
\end{IEEEbiography}

\vskip -1\baselineskip
\begin{IEEEbiography}
[{\includegraphics[width=1in,height=1.25in,clip,keepaspectratio]{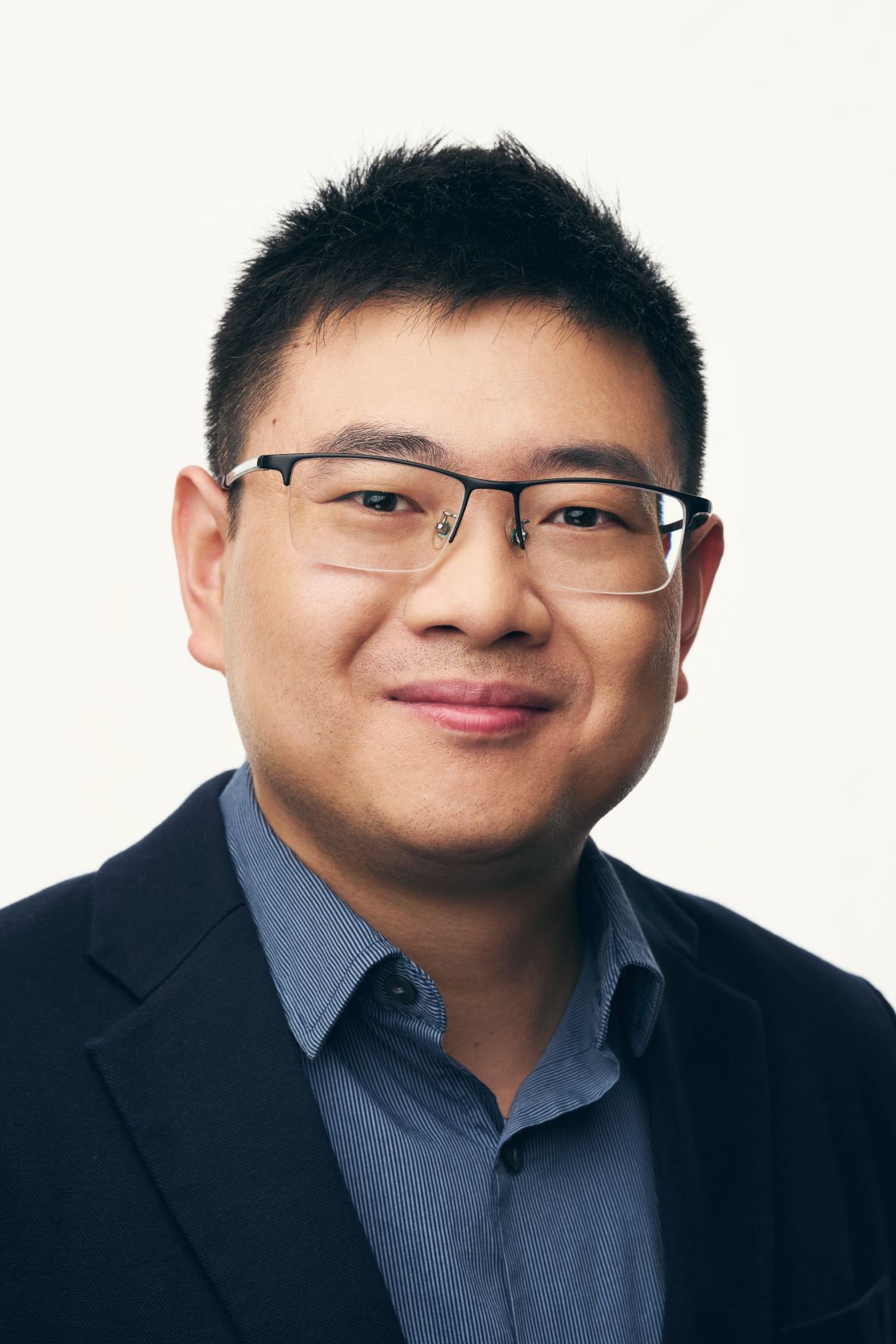}}]{Lei S. Li} is an assistant professor of electrical and computer engineering and bioengineering at Rice University. He obtained his PhD from the Department of Electrical Engineering at the California Institute of Technology in 2019. He received his MS degree at Washington University in St. Louis in 2016. His research focuses on developing next-generation medical imaging technology for understanding the brain better, diagnosing early-stage cancer, and wearable monitoring of human vital signs.
\end{IEEEbiography}

\vskip -1\baselineskip
\begin{IEEEbiography}
[{\includegraphics[width=1in,height=1.25in,clip,keepaspectratio]{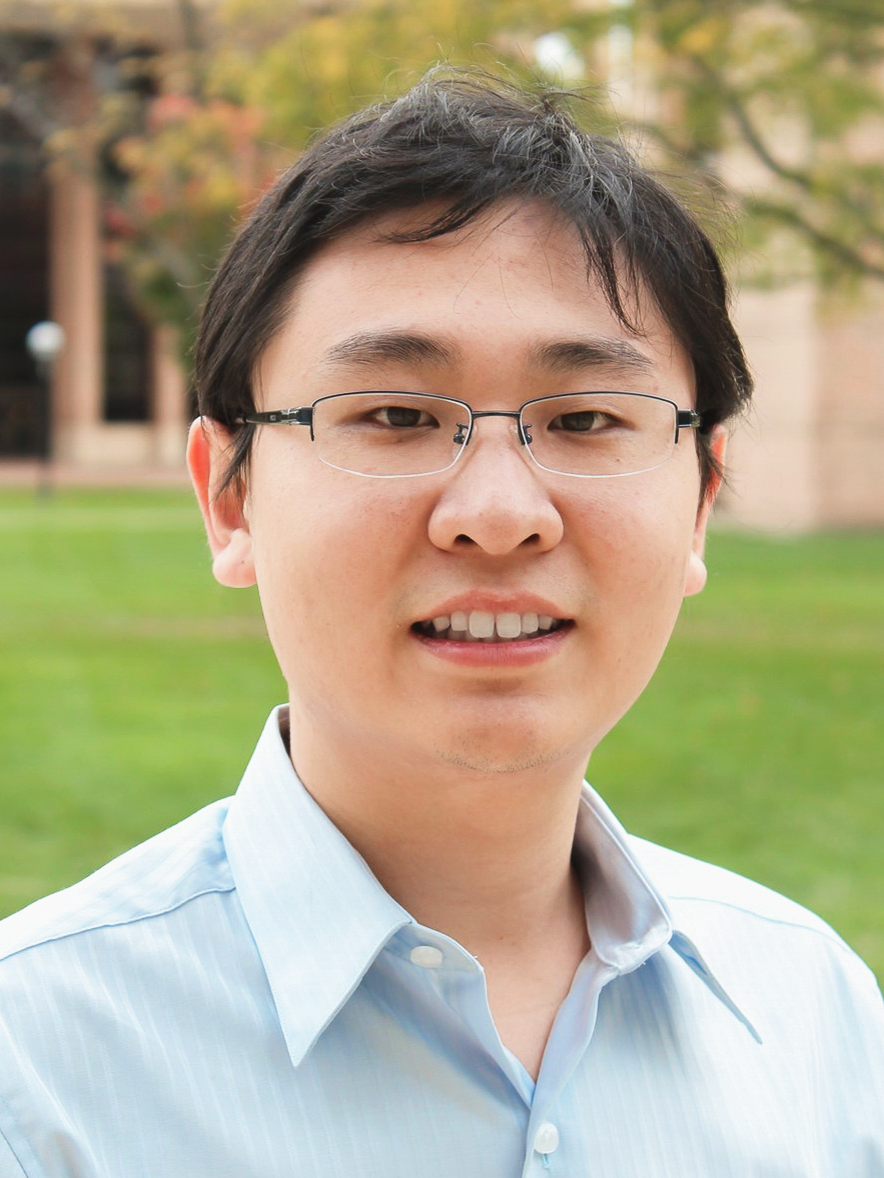}}] {Kaiyuan Yang} (Senior Member, IEEE) is an Associate Professor of Electrical and Computer Engineering at Rice University, USA, where he leads the Secure and Intelligent Micro-Systems (SIMS) lab. He received a B.S. degree in Electronic Engineering from Tsinghua University, China, in 2012, and a Ph.D. degree in Electrical Engineering from the University of Michigan - Ann Arbor, MI, in 2017. His research focuses on low-power integrated circuits and system design for bioelectronics, hardware security, and mixed-signal/in-memory computing. 

Dr. Yang is a recipient of NSF CAREER Award, IEEE SSCS New Frontier Award, SSCS Predoctoral Achievement Award, and best paper awards from premier conferences in multiple fields, including 2024 Annual International Conference of the IEEE Engineering in Medicine and Biology Society (EMBC), 2022 ACM Annual International Conference on Mobile Computing and Networking (MobiCom), 2021 IEEE Custom Integrated Circuit Conference (CICC), 2016 IEEE International Symposium on Security and Privacy (Oakland), and 2015 IEEE International Symposium on Circuits and Systems (ISCAS). His research was also selected as the research highlights of Communications of ACM and ACM GetMobile magazines, and IEEE Top Picks in Hardware and Embedded Security. He currently serves as an associate editor of IEEE Transactions on VLSI Systems (TVLSI) and a program committee member of ISSCC, CICC, ISCA, and MICRO conferences. 
\end{IEEEbiography}

\vspace{11pt}

\vspace{11pt}

\vfill

\end{document}